\documentclass[journal]{IEEEtran}
\usepackage{amssymb}
\usepackage{amsbsy}
\usepackage{amsmath}
\usepackage{amsthm} 
\usepackage{setspace}
\usepackage{epsfig}
\usepackage{cite}
\usepackage{graphics}
\usepackage{epstopdf}
\usepackage{mathrsfs}
\usepackage[above]{placeins}
\usepackage{algorithm}
\usepackage{algpseudocode}
\usepackage{multirow}
\usepackage{algorithmicx}
\usepackage{tikz}

\begin{document}
\title{\huge{Sphere Decoding Revisited}}

\author{Zheng~Wang,~\IEEEmembership{Senior Member, IEEE,}
        Cong~Ling,~\IEEEmembership{Member, IEEE,}
        Shi~Jin,~\IEEEmembership{Fellow, IEEE,}
        Yongming~Huang,~\IEEEmembership{Fellow, IEEE}
        and Feifei~Gao,~\IEEEmembership{Fellow, IEEE}
\thanks{

Z. Wang, S. Jin and Y. Huang are with National Mobile Communications Research Laboratory, and School of Information Science and Engineering, Southeast University, Nanjing 210096, China (e-mail: wznuaa@gmail.com);
C. Ling is with the Department of Electrical and Electronic Engineering, Imperial College London, London, SW7 2AZ, United Kingdom; F. Gao is with the Department of Automation, Tsinghua University, Beijing 100084, China.
}}


\maketitle

\begin{abstract}
In this paper, the paradigm of sphere decoding (SD) for solving the integer least square problem (ILS) is revisited, where extra degrees of freedom are introduced to exploit the decoding potential.
Firstly, the equivalent sphere decoding (ESD) is proposed, which is essentially the same as the classic Fincke-Pohst sphere decoding but characterizes the sphere radius $D>0$ with two new parameters named as initial searching size $K>1$ and deviation factor $\sigma>0$.
By fixing $\sigma$ properly, we show that given the sphere radius $D\triangleq\sigma\sqrt{2\ln K}$, the complexity of ESD in terms of the number of visited nodes is upper bounded by $|S|<nK$, thus resulting in an explicit and tractable decoding trade-off solely controlled by $K$. To the best of our knowledge, this is the first time that the complexity of sphere decoding is exactly specified, where considerable decoding potential can be explored from it.
After that, two enhancement mechanisms named as normalized weighting and candidate protection are proposed to further upgrade the ESD algorithm. On one hand, given the same setups of $K$ and $\sigma$, a larger sphere radius is achieved, indicating a better decoding trade-off. On the other hand, the proposed ESD algorithm is generalized, which bridges suboptimal and optimal decoding performance through the flexible choice of $K$.
Finally, further performance optimization and complexity reduction with respect to ESD are also derived, and the introduced tractable and flexible decoding trade-off is verified through large-scale MIMO detection.

\end{abstract}

\IEEEpeerreviewmaketitle


\IEEEpeerreviewmaketitle


\section{Introduction}
\IEEEPARstart{N}{o}wadays, the large-scale multiple-input multiple-output (MIMO) system has become a promising extension of MIMO in 5G, which boosts the network capacity on a much greater scale without extra bandwidth \cite{5595728,6736761,6375940,6457363}. However, the dramatically increased system size also places a pressing challenge on signal detection, which belongs to integer least square (ILS) problem.
As a general way to realize the maximum-likelihood (ML) decoding, the traditional sphere decoding (SD) turns out to be impractical due to the unaffordable complexity in large-scale systems \cite{Kannan,Agrell2002,HassibiExpected,DamenDetectionSearch}.
As for those near-ML decoding schemes like fixed-complexity sphere decoding (FCSD), K-best decoder, etc., they are also inapplicable
due to the intensive complexity increment and terrible performance deterioration \cite{ZhengWangTSP2019,Hassibisoft,JaldenFSD,4154776,EmbeddingLuzzi}.
To this end, most of related works try to focus on the low-complexity decoding schemes although their performance is severely limited \cite{TabuSrinidhi,DaiL1,GaoX1,MSMIMO1,6008619,ZhengWangTCOM2019}.



Theoretically, a fundamental problem in the framework of sphere decoding is the inexplicit decoding trade-off between performance and complexity.
Even though it is well known that sphere decoding entails the exponentially increasing complexity with the increment of the system dimension, the relationship between complexity and performance has not been well revealed.
Take the classic Fincke-Pohst SD (which is the same as the sphere decoding presented by Hassibi in \cite{HassibiExpected}) as an example, it is easy to set the sphere radius $D>0$ freely to determine the decoding performance but the corresponding decoding complexity (e.g., the number of visited nodes during the searching process $|S|$) cannot be specified \cite{FPSD}. This heavily restricts the development of sphere decoding especially in high-dimensional systems.
In \cite{HassibiExpected}, an average version of $|S|$ for Fincke-Pohst SD was derived, which was further improved through the analysis of its asymptotic behaviour in \cite{JaldenSphere}. However, they mainly focused on characterizing the mean and variance of the complexity for i.i.d. Gaussian lattice basis. In \cite{6006604}, the tail exponents of the SD complexity distribution were investigated for the complexity estimation. Nevertheless, the number of visited nodes $|S|$ in sphere decoding is evaluated in a probabilistic manner.
Other works about complexity of SD can be found in \cite{6216420,5429127}, which either takes the specific conditions from communications into account or considers the complexity in infinity-norm SD.

%


In this paper, we try to answer this fundamental problem of sphere decoding by giving the clear relationship between sphere radius $D$ and the number of visited nodes $|S|$, thus enabling a tractable and flexible decoding trade-off between performance and complexity. Based on this new paradigm of sphere decoding, remarkable potential can be exploited to make it competitive even under high-dimensional systems.
To summarize, we advance the state of the art of sphere decoding in the following several fronts:

First of all, \emph{equivalent sphere decoding} (ESD) is proposed to solve the ILS problem (i.e., CVP in lattice decoding) in large-scale MIMO detection, which introduces two new parameters named as \emph{initial searching size} $K>1$ and \emph{deviation factor} $\sigma>0$ to characterize the concept of sphere radius, i.e.,
$D\triangleq\sigma\sqrt{2\ln K}$.
Intuitively, compared to the traditional sphere decoding, extra degrees of freedom are obtained in interpreting the decoding trade-off.
In particular, by fixing $\sigma$ reasonably, we show that given the sphere radius $D=\sigma\sqrt{2\ln K}$, the number of visited nodes during the searching process is upper bounded by $|S|<nK$, where $n$ is the system dimension.
Since ESD actually works the same as Fincke-Pohst SD but with an explicit and tractable decoding trade-off between performance and complexity, the afore-mentioned problem about sphere decoding is addressed.

Secondly, two mechanisms referred to as \emph{normalized weighting} and \emph{candidate protection} are introduced to upgrade the proposed ESD algorithm. Thanks to normalized weighting, ESD is able to achieve a larger sphere radius under the same sizes of $K$ and $\sigma$, which indicates a better tractable decoding trade-off. On the other hand, because of candidate protection, suboptimal decoding solutions still can be outputted by ESD for the given small sizes of $K$. This leads to a flexible decoding ranging from suboptimal performance (i.e., $K=1$ corresponds to Babai's nearest plane algorithm) to optimal ML performance.
We emphasize that such a generalization from ML decoding to bounded distance decoding (BDD) is rather crucial for the development of SD. As an ML decoding scheme, sphere decoding has been ignored for a long time due to the rise of decoding large-scale problems.
More interestingly, we demonstrate that even with the enhancements by normalized weighting and candidate protection, the complexity of ESD in terms of the number of visited nodes still follows the upper bounded $|S|<nK$.
Therefore, the proposed ESD algorithm establishes a novel framework with respect to sphere decoding, which suits the high-dimensional systems well by its promising tractability, flexibility and efficiency.

Thirdly, further performance optimization and complexity reduction regarding to the proposed ESD algorithm are investigated to make it more competitive.
Meanwhile, from the point of view of lattice Gaussian distribution, the perspective decoding potential of ESD is also studied.
The implementation of ESD with finite state space is carefully investigated under the consideration of decoding efficiency.
Given the initial searching size $K$, the choice of deviation $\sigma$ is optimized by relaxation. In addition, Lenstra-Lenstra-Lov\'{a}sz (LLL) reduction from lattice decoding is applied to serve as a preprocessing stage for ESD, and we show that the related sphere radius under the help of LLL reduction would be significantly improved with polynomial complexity $O(n^3\log n)$.

The rest of this paper is organized as follows. Section II introduces the system model and briefly reviews the basics of sphere decoding and Babai's nearest plane algorithm.
In Section III, equivalent sphere decoding (ESD) algorithm is proposed, followed by the related analysis in both decoding performance and complexity.
In Section IV and V, normalized weighting and candidate protection are proposed for ESD respectively to improve the decoding trade-off and enable flexibility.
In Section VI, further performance optimization and complexity reduction are provided, and simulation results via large-scale MIMO detection are presented in Section VII. Finally, Section VIII concludes the paper.


\emph{Notation:} Matrices and column vectors are denoted by upper
and lowercase boldface letters, and the transpose, inverse, pseudoinverse
of a matrix $\mathbf{B}$ by $\mathbf{B}^T, \mathbf{B}^{-1},$ and
$\mathbf{B}^{\dag}$, respectively. We use $\mathbf{b}_i$ for the $i$th
column of the matrix $\mathbf{B}$, $b_{i,j}$ for the entry in the $i$th row
and $j$th column of the matrix $\mathbf{B}$. $\lceil x \rfloor$ denotes rounding to
the integer closest to $x$. If $x$ is a complex number, $\lceil x \rfloor$
rounds the real and imaginary parts separately.
Finally, in this paper, the complexity of SD is evaluated by the number of visited nodes (i.e., $|S|$) during the searching along the tree traversal.
Meanwhile, the computational complexity is measured by the number of arithmetic operations (additions, multiplications, comparisons, etc.).

\newtheorem{my1}{Lemma}
\newtheorem{my2}{Theorem}
\newtheorem{my3}{Definition}
\newtheorem{my4}{Proposition}
\newtheorem{my5}{Remark}
\newtheorem{my6}{Conjection}
\newtheorem{my7}{Corollary}

\section{Preliminaries}
In this section, besides the system model, we also introduce the background and the mathematical tools needed to describe and analyze the proposed ESD algorithm.

\subsection{System Model}
Given the full $n\times m$ column-rank matrix $\mathbf{B}\in\mathbb{R}^{n\times m}$ with $n\geq m$, the $n$-dimensional lattice $\Lambda$ generated by it is defined as
\begin{equation}
\Lambda=\{\mathbf{Bx}: \mathbf{x}\in \mathbb{Z}^m\},
\end{equation}
where $\mathbf{B}$ is called the lattice basis.
Here, for notational simplicity, we assume $n=m$ throughout the context and consider the decoding of an $n \times n$ real-valued system. The extension to the complex-valued system is straightforward \cite{CongRandom,Xia1}.
Then, let $\mathbf{x}\in \mathbb{Z}^n$ denote the transmitted signal, the corresponding received signal $\mathbf{c}$ over MIMO systems is given by
\begin{equation}
\mathbf{c}=\mathbf{B}\mathbf{x}+\mathbf{w}
\label{eqn:System Model}
\end{equation}
where $\mathbf{w}\in \mathbb{R}^n$ is the noise vector with zero mean and variance $\sigma_w^{2}$.
Typically, the conventional maximum likelihood (ML) decoding reads
\begin{equation}
\widehat{\mathbf{x}}_{\text{ML}}=\underset{\mathbf{x}\in \mathbb{Z}^{n}}{\operatorname{arg~min}} \, \|\mathbf{B}\mathbf{x}-\mathbf{c}\|^2
\label{eqn:ML Decoding}
\end{equation}
where $\|\cdot\|$ denotes the Euclidean norm. Clearly, the ML decoding in the above MIMO systems corresponds to the integer least square (ILS) problem, which is also known as the closest vector problem (CVP) in lattice theory \cite{DamenDetectionSearch}.



\subsection{Fincke-Pohst Sphere Decoding}
For a better presentation, the QR-decomposition with $\mathbf{B}=\mathbf{QR}$ is applied, and the system model in (\ref{eqn:System Model}) can be expressed as
\begin{equation}
\mathbf{y}=\mathbf{Q}^T\mathbf{c}=\mathbf{R}\mathbf{x}+\mathbf{n},
\label{eqn:QR}
\end{equation}
where $\mathbf{Q}\in\mathbb{R}^{n\times n}$ is an orthogonal matrix and $\mathbf{R}\in\mathbb{R}^{n\times n}$ is an upper triangular matrix. Accordingly, the ML decoding in (\ref{eqn:ML Decoding}) becomes
\begin{equation}
\widehat{\mathbf{x}}_{\text{ML}}=\underset{\mathbf{x}\in \mathbb{Z}^{n}}{\operatorname{arg~min}} \, \|\mathbf{R}\mathbf{x}-\mathbf{y}\|^2.
\label{eqn:ML DecodingQR}
\end{equation}


To address the ILS problem in (\ref{eqn:ML DecodingQR}), in \emph{Babai's nearest plane} algorithm, $\widehat{x}_i$ is decoded in a backwards order layer by layer (i.e., $i=n,n-1,\ldots,1$) through direct rounding \cite{Babai}
\begin{equation}
\widehat{x}_i=\lceil\widetilde{x}_i\rfloor,
\label{rounding}
\end{equation}
where
\begin{equation}
\widetilde{x}_i=\frac{y_i-\sum^n_{j=i+1}r_{i,j}\widehat{x}_j}{r_{i,i}}.
\label{eqn:noise effection in determination}
\end{equation}
In this way, the latent interference from elements $x_{i+1}, \ldots, x_{n}$ of $\mathbf{x}$ can be alleviated for the decoding of $x_i$.
Babai's nearest plane algorithm is also referred to as \emph{successive interference cancellation} (SIC) algorithm in the field of MIMO detection \cite{6182560}.





\begin{figure}[t]
\begin{center}
\includegraphics[width=3.2in]{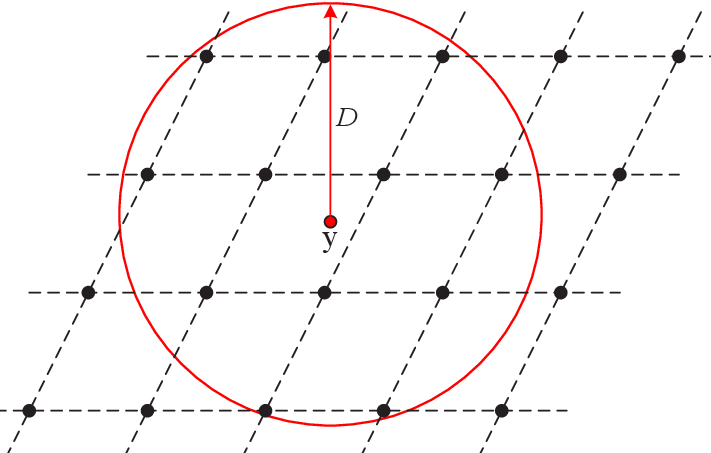}
\end{center}
\vspace{-1em}
  \caption{The illustration of sphere decoding with respect to a 2-dimensional system model in (\ref{sd1}).}
  \label{simulation x}
\end{figure}

On the other hand, to achieve ML decoding, the classic Fincke-Pohst SD was applied to enumerate all the possible lattice points $\mathbf{Rx}$ within a sphere radius $D>0$ \cite{FPSD,HassibiExpected}:
\begin{equation}
\|\mathbf{Rx}-\mathbf{y}\|\leq D.
\label{sd1}
\end{equation}

Specifically, based on the upper triangular matrix $\mathbf{R}$, (\ref{sd1}) can be further expressed as
\begin{equation}
D^2\geq\sum_{i=1}^n\left(y_i-\sum_{j=i}^nr_{i,j}x_j\right)^2.
\end{equation}
Then, considering the searching at the first layer i.e., $i=n$, the above restriction by enumeration becomes
\begin{equation}
D^2\geq (y_n-r_{n,n}x_n)^2,
\end{equation}
where the searching space of $\widehat{x}_n$ belongs to the interval
\begin{equation}
\left\lceil\frac{D+y_n}{r_{n,n}}\right\rceil\leq\widehat{x}_n\leq\left\lfloor\frac{D+y_n}{r_{n,n}}\right\rfloor.
\end{equation}


Let $\widehat{x}^j_i$ denote the $j$th closest integer candidate node to $\widetilde{x}_i$,
the searching space of $\widehat{x}_i$ in the recursive decoding from layer $n$ to $1$ can be written as \cite{HassibiExpected}
\begin{equation}
|\widehat{x}_i\hspace{-.2em}-\hspace{-.2em}\widetilde{x}_i|_{\text{Fincke-Pohst}}\hspace{-.2em}\leq \hspace{-.3em}\sqrt{D^2\hspace{-.2em}-\hspace{-.3em}\sum_{j=i+1}^n \left| y_j\hspace{-.2em}-\hspace{-.2em}\sum_{l=j}^n r_{j,l}\widehat{x}_l \right|^2}/|r_{i,i}|,
\label{sdsame}
\end{equation}
where candidate node $\widehat{x}^j_i$ at layer $i$ satisfying (\ref{sdsame}) will be saved.
In this way, the searching space is expanded layer by layer like a tree until $i=1$ while candidate vectors $\widehat{\mathbf{x}}$'s made up from candidate nodes are obtained through this tree-search decoding structure.

Consequently, among all the collected candidate vectors $\widehat{\mathbf{x}}$'s within the sphere radius $D$, the one with the smallest Euclidean distance $\|\mathbf{R\widehat{x}}-\mathbf{y}\|$ will be outputted as the final decoding solution.
Note that the sphere radius $D$ has to be selected carefully since a large one would lead to considerable complexity waste while no eligible candidate vectors would be yielded with a small choice of $D$ \cite{Phost}.
However, since sphere decoding was introduced, there has been a fundamental problem as its decoding trade-off between performance and complexity is not clear enough, which severely limits its development in these years especially for high-dimensional systems.


\renewcommand{\algorithmicrequire}{\textbf{Input:}}  
\renewcommand{\algorithmicensure}{\textbf{Output:}} 

\section{Equivalent Sphere Decoding Algorithm}
In this section, equivalent sphere decoding (ESD) algorithm is proposed to reveal the explicit decoding trade-off between the sphere radius and the number of visited nodes.

\subsection{Algorithm Description}
First of all, the recursive searching layer by layer in Fincke-Pohst SD is retained by ESD and the searching is still performed layer by layer in a backwards order from $i=n$ to $i=1$.
To concisely state the operations, the following definitions based on the tree-search structure are made.


Define the \emph{initial pruning size} $K>1, K\in\mathbb{R}$, which is set initially to control the algorithm performance and complexity.
Accordingly, the \emph{searching size} $K(\widehat{x}^j_i)>0, K(\widehat{x}^j_i)\in\mathbb{R}$ for each integer candidate node $\widehat{x}^j_i$ is defined
as
\begin{equation}
K(\widehat{x}^j_i)\triangleq K(\underline{\widehat{x}^j_{i}})\cdot f(\widehat{x}^j_i)
\label{prunethresholdla}
\end{equation}
with defined \emph{weighting} function
\begin{equation}
f(\widehat{x}^j_i)\triangleq e^{-\frac{1}{2\sigma^2_i}\|\widehat{x}^j_i-\widetilde{x}_i\|^2},
\label{fn1}
\end{equation}
where $\sigma_i\triangleq\frac{\sigma}{|r_{i,i}|}$, and $\sigma>0$ denotes the \emph{deviation factor} to adjust the weighting $f(\cdot)$.
Here, $\underline{\widehat{x}^j_{i}}$ indicates the parent node of $\widehat{x}^j_{i}$ at the previous searching layer $i+1$. It is easy to check that the initial searching size $K=K(\underline{\widehat{x}^j_n})$.
Note that several children candidate nodes $\widehat{x}^{j}_{i}$ may have a same parent node $\underline{\widehat{x}^j_{i}}$.


Next, based on the searching size $K(\widehat{x}^j_i)$, the integer candidate node $\widehat{x}^j_i$ at layer $i$ will be saved if it satisfies the following \emph{pruning threshold}
\begin{equation}
K(\widehat{x}^j_i)\geq 1.
\label{prunethreshold1b}
\end{equation}
Otherwise, the candidate node $\widehat{x}^j_i$ will be pruned while the searching steps into the next layer $i-1$ given those saved candidate nodes.
As shown in Fig. 2, candidate nodes $\widehat{x}^1_{i-1}$ and $\widehat{x}^2_{i-1}$ at layer $i-1$ are saved to enable the searching at the next layer.
Intuitively, since the weighting $f(\widehat{x}^j_i)$ is exponentially decayed with the index $j$, all the candidate nodes with index $j>3$ are deterministically pruned if node $\widehat{x}^3_{i-1}$ fails to satisfy the pruning threshold.

\begin{figure}[h]
\begin{center}
\hspace{-1em}\includegraphics[width=2.6in]{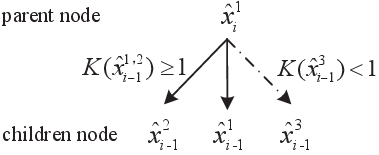}
\end{center}
\vspace{-1em}
  \caption{The illustration of pruning threshold, where candidate nodes $\widehat{x}^3_{i-1}$ and $\widehat{x}^{j>3}_{i-1}$ stemmed from $\widehat{x}^1_{i}$ are pruned.}
  \label{simulation x1}
\end{figure}

According to the proposed pruning threshold, the searching along the tree structure works layer by layer while the survived decoding candidate vectors (i.e., $\widehat{\mathbf{x}}=[x_1,\ldots,x_n]^T$) are saved by a candidate list $L$.
Finally, the candidate vector $\widehat{\mathbf{x}}$ with the smallest Euclidean distance $\|\mathbf{R\widehat{x}}-\mathbf{y}\|$ among the candidate list $L$ will be outputted as the decoding solution.

\begin{my1}
Given the initial searching size $K>1$, candidate vectors $\mathbf{x}$'s within sphere radius
\end{my1}
\vspace{-1.5em}
\begin{equation}
\|\mathbf{Rx}-\mathbf{y}\|\leq \sigma\sqrt{2\ln K}
\label{b12a}
\end{equation}
\emph{will be obtained by ESD.}
\begin{proof}
According to (\ref{prunethresholdla}) and (\ref{prunethreshold1b}), the pruning threshold can be further expressed inductively by
\begin{equation}
f(\widehat{x}^j_i)\geq\frac{1}{K(\underline{\widehat{x}^j_i})}=\frac{1}{K\cdot f(\widehat{x}^j_{i+1})\cdots f(\widehat{x}^j_{n})}.
\label{b12b}
\end{equation}

On the other hand, (\ref{b12a}) can be rewritten as
\begin{equation}
e^{-\frac{1}{2\sigma^2}\|\mathbf{Rx}-\mathbf{y}\|^2}\geq \frac{1}{K},
\label{b12c}
\end{equation}
which can be further expressed by factorization as
\begin{equation}
\prod^n_{i=1}e^{-\frac{1}{2\sigma^2_{n-i+1}}\|\widehat{x}_{n-i+1}-\widetilde{x}_{n-i+1}\|^2}\hspace{-.5em}=\prod^n_{i=1}f(\widehat{x}_{n-i+1})\geq \frac{1}{K}.
\label{b12d}
\end{equation}

Hence, in order to collect candidate vectors that satisfy (\ref{b12d}), considering the fact that $0<f(\cdot)\leq1$, $f(\widehat{x}_i)$ should fulfill the following requirement
{\allowdisplaybreaks\begin{flalign}
f(\widehat{x}_i)&\geq\frac{1}{K\cdot\prod_{j\neq i}f(\widehat{x}_j)}\notag\\
&\geq\frac{1}{K\cdot f(\widehat{x}_{i+1})\cdots f(\widehat{x}_{n})}
\end{flalign}}for $1\leq i\leq n$, which exactly corresponds to the proposed pruning threshold in (\ref{b12b}).
\end{proof}


From (\ref{b12b}), the initial searching size $K$ essentially serves as a parameter to adjust the pruning threshold, thus determining the related sphere radius.
Intuitively, a larger size $K$ corresponds to a smaller pruning threshold in (\ref{b12b}) and more visited nodes at each layer, thereby saving more decoding candidate vectors in the end.

Based on Lemma 1, we now verify the equivalence of Fincke-Pohst SD and ESD by showing they have the same searching space of $\widehat{x}_i$ at each layer.
\begin{my2}
By sharing the same searching space at each layer, the proposed ESD is exactly the same as Fincke-Pohst SD with sphere radius
\end{my2}
\vspace{-1em}
\begin{equation}
D=\sigma\sqrt{2\ln K}.
\label{mp11}
\end{equation}
\begin{proof}
According to the pruning threshold in (\ref{b12b}), the searching space of $\widehat{x}_i$ given $\widetilde{x}_i$ in ESD can be derived as
{\allowdisplaybreaks\begin{flalign}
|\widehat{x}_i-\widetilde{x}_i|_{\text{ESD}}&\leq \hspace{-0.3em} \sqrt{2\sigma^2\ln K-\sum_{j=i+1}^n \left| y_j-\sum_{l=j}^n r_{j,l}\widehat{x}_l \right|^2}/|r_{i,i}|\notag\\
&= \hspace{-0.3em}\sqrt{D^2-\sum_{j=i+1}^n \left| y_j-\sum_{l=j}^n r_{j,l}\widehat{x}_l \right|^2}/|r_{i,i}|\notag\\
&= |\widehat{x}_i-\widetilde{x}_i|_{\text{Fincke-Pohst}},
\end{flalign}}which is exactly the boundary of Fincke-Pohst SD in (\ref{sdsame}) for $1\leq i\leq n$.
\end{proof}


Different from the paradigm of conventional sphere decoding in (\ref{sd1}), the optimization paradigm in ESD for solving the ILS problem is transformed into (\ref{b12c}).
More specifically, it is easy to verify that
\begin{equation}
\widehat{\mathbf{x}}_{\text{ML}}=\underset{\mathbf{x}\in \mathbb{Z}^{n}}{\operatorname{arg~min}} \, \|\mathbf{R}\mathbf{x}-\mathbf{y}\|^2=\underset{\mathbf{x}\in \mathbb{Z}^{n}}{\operatorname{arg~max}}\ e^{-\frac{1}{2\sigma^2}\|\mathbf{R}\mathbf{x}-\mathbf{y}\|^2}.
\end{equation}
Although ESD works the same as Fincke-Pohst SD, along the searching process it takes advantages of two parameters $K$ and $\sigma$ rather than the single one $D$, which offers more degrees of freedom to deal with the ILS problem.
Typically, it is clear to see that the above equivalence always holds no matter what the deviation factor $\sigma>0$ is, which indicates $\sigma$ should be carefully selected for a better decoding.





\subsection{Explicit Trade-off Between Performance and Complexity}
According to $D=\sigma\sqrt{2\ln K}$ in Theorem 1, extra freedom can be obtained in interpreting the operations of sphere decoding, which provides a feasible way for the analytical diagnosis in both decoding performance and complexity.

\begin{my1}
In ESD, for each parent candidate node $\underline{\widehat{x}^j_{i}}$ with $K(\underline{\widehat{x}^j_i})\geq1$,
the number of its saved children candidate nodes at decoding layer $i$ satisfies
\end{my1}
\vspace{-1em}
\begin{equation}
K_{\text{save}}\leq K(\underline{\widehat{x}^j_i})
\label{pj1ab}
\end{equation}
if $\sigma<\min_i|r_{i,i}|/(2\sqrt{2\ln 2})$.
\begin{proof}
According to the pruning threshold given in (\ref{prunethreshold1b}), the condition shown in (\ref{pj1ab}) holds if and only if the $\lfloor K(\underline{\widehat{x}^j_i})+1\rfloor$th closest integer candidate to $\widetilde{x}_i$ will definitely be pruned, that is
\begin{equation}
K(\underline{\widehat{x}^j_i})f(\widehat{x}^{\lfloor K(\underline{\widehat{x}^j_i})+1 \rfloor}_i)< 1.
\label{wo1a}
\end{equation}

Then, because the distance $|\widehat{x}^j_i-\widetilde{x}_i|$ is bounded by
\begin{equation}
(j-1)\cdot\frac{1}{2}\leq|\widehat{x}^j_i-\widetilde{x}_i|\leq j\cdot\frac{1}{2},
\label{wo31}
\end{equation}
(\ref{wo1a}) can be achieved if
\begin{equation}
K(\underline{\widehat{x}^j_i})\cdot e^{-\frac{1}{8\sigma_i^2}(\lfloor K(\underline{\widehat{x}^j_i})+1\rfloor-1)^2}<1,
\label{wo4a}
\end{equation}
which corresponds to
\begin{equation}
\sigma^2<\frac{(\lfloor K(\underline{\widehat{x}^j_i})+1\rfloor-1)^2}{8\ln K(\underline{\widehat{x}^j_i})}\cdot r^2_{i,i}.
\label{wo5al}
\end{equation}

Moreover, it is easy to confirm the lower bound of the right-hand side (RHS) of (\ref{wo5al}) as
\begin{equation}
\frac{(\lfloor K(\underline{\widehat{x}^j_i})+1\rfloor-1)^2}{8\ln K(\underline{\widehat{x}^j_i})}\cdot r^2_{i,i}>\frac{1}{8\ln 2}\cdot r^2_{i,i},
\label{wo5ab}
\end{equation}
which means (\ref{wo5al}) is fulfilled if
\begin{equation}
\sigma<\min_i |r_{i,i}|/(2\sqrt{2\ln 2})
\end{equation}
for $1\leq i\leq n$.
\end{proof}

\begin{my1}
In ESD, for each parent candidate node $\underline{\widehat{x}^j_{i}}$ with $K(\underline{\widehat{x}^j_i})\geq1$,
the summation of searching sizes of its saved children candidate nodes at decoding layer $i$ is decreasing
\end{my1}
\vspace{-1em}
\begin{equation}
\sum_jK(\widehat{x}^j_i)< K(\underline{\widehat{x}^j_i})
\label{pj1a1}
\end{equation}
if $\sigma\leq\min_{i}|r_{i,i}|/(2\sqrt{\pi})$.
\begin{proof}
Based on the definition given in (\ref{prunethresholdla}), it follows that
{\allowdisplaybreaks\begin{flalign}
\sum_jK(\widehat{x}^j_i)&=K(\underline{\widehat{x}^j_i})\cdot\sum_jf(\widehat{x}^j_i)\notag\\
&<K(\underline{\widehat{x}^j_i})\cdot \sum_{\widehat{x}_i\in\mathbb{Z}}e^{-\frac{1}{2\sigma^2_i}\|\widehat{x}_i-\widetilde{x}_i\|^2}\notag\\
&\overset{(a)}{\leq} K(\underline{\widehat{x}^j_i})\cdot \sum_{\widehat{x}_i\in\mathbb{Z}}e^{-\frac{1}{2\sigma^2_i}\|\widehat{x}_i\|^2}\notag\\
&\overset{(b)}{=}K(\underline{\widehat{x}^j_i})\cdot \vartheta_3(|r_{i,i}|^2/2\pi\sigma^2)\notag\\
&\overset{(c)}{\approx}K(\underline{\widehat{x}^j_i}).
\label{pj1a1}
\end{flalign}}Here, inequality (a) recalls the following relationship
\begin{equation}
\rho_{\sigma,\mathbf{y}}(\Lambda)\leq\rho_{\sigma}(\Lambda),
\label{v1f}
\end{equation}
where $\rho_{\sigma, \mathbf{y}}(\Lambda)= \sum_{\mathbf{x} \in \mathbb{Z}^n}e^{-\frac{1}{2\sigma^2}\parallel \mathbf{Rx}-\mathbf{y} \parallel^2}$, $\rho_{\sigma}(\Lambda)= \sum_{\mathbf{x} \in \mathbb{Z}^n}e^{-\frac{1}{2\sigma^2}\parallel \mathbf{Rx} \parallel^2}$ and the equality holds only when $\mathbf{y}\in\Lambda$ \cite{MicciancioGaussian}. Inequality (b) invokes the \emph{Jacobi theta function} $\vartheta_3$ \cite{ConwayandSloane}
\begin{equation}
\vartheta_3(\nu)=\sum^{+\infty}_{n=-\infty}e^{-\pi\nu n^2},
\label{b16}
\end{equation}
and the approximation in (c) follows
\begin{equation}
\prod_{i=1}\vartheta_3(|r_{i,i}|^2/2\pi\sigma^2)\leq\vartheta_3(2)=1.0039\approx1
\label{bound7}
\end{equation}
for $\sigma\leq\min_{i}|r_{i,i}|/(2\sqrt{\pi})$ because $\vartheta_3(\nu)$ is monotonically decreasing with $\nu>0$.
\end{proof}

Based on Lemmas 2 \& 3, the complexity of ESD can be derived by means of the number of visited nodes as follows.
\begin{my2}
In ESD, let $\sigma=\min_{i}|r_{i,i}|/(2\sqrt{\pi})$, the number of visited nodes denoted by $|S|$ is upper bounded by
\end{my2}
\vspace{-1em}
\begin{equation}
|S|< nK.
\end{equation}
\begin{proof}
According to Lemma 2, the number of saved candidate nodes at each searching layer is upper bounded by the summation of searching sizes at the previous layer, i.e.,
\begin{equation}
K^{\text{layer}\ i}_{\text{save}}=\sum K_{\text{save}}\leq \sum K(\underline{\widehat{x}^j_i})=K^{\text{layer}\ i+1}_{\text{search\ size}}.
\end{equation}
Meanwhile, from Lemma 3, because the summation of searching sizes at each searching layer is decreasing as
\begin{equation}
K^{\text{layer}\ 1}_{\text{search\ size}}<\ldots< K^{\text{layer}\ n}_{\text{search\ size}}< K^{\text{layer}\ n+1}_{\text{search\ size}}=K
\label{pj1pa}
\end{equation}
so that the number of visited nodes is upper bounded by
\begin{equation}
|S|=\sum^n_{i=1}K^{\text{layer}\ i}_{\text{save}}\leq \sum^n_{i=1}K^{\text{layer}\ i+1}_{\text{search\ size}}< nK.
\end{equation}
This completes the proof.
\end{proof}


Theorem 2 is rather crucial to the study of sphere decoding as the complexity of sphere decoding can be specified through the upper bound for the first time.
To this end, one can simply fix $\sigma$ and enjoy the decoding trade-off through the single tunable parameter $K$, which naturally leads to the following Corollary.
\begin{my7}
With $\sigma=\min_{i}|r_{i,i}|/(2\sqrt{\pi})$, ESD achieves the tractable sphere radius $D=\sqrt{\frac{\ln K}{2\pi}}\min_{i}|r_{i,i}|$ with the number of visited nodes upper bounded by $|S|< nK$.
\end{my7}

From Corollary 1, we know that in order to increase the sphere radius $D$, $K$ should increase exponentially, which corresponds to an exponentially increased complexity of sphere decoding.
This is in accordance with the common sense of sphere decoding but with a specified complexity upper bound.
On the other hand, because the number of saved candidate nodes at searching layer $i=1$ accounts for the number of collected candidate vectors i.e., $K^{\text{layer}\ 1}_{\text{save}}=|L|$, we can easily arrive at the following result.
\begin{my7}
With $\sigma=\min_{i}|r_{i,i}|/(2\sqrt{\pi})$, the number of candidate vectors collected by ESD denoted by $|L|$ is upper bounded by
\end{my7}
\vspace{-1em}
\begin{equation}
|L|< K.
\end{equation}

\begin{algorithm}[t]
\caption{Equivalent Sphere Decoding (ESD)}
\begin{algorithmic}[1]
\Require
$K, \mathbf{R}, \mathbf{y}, \sigma=\min_{i}|r_{i,i}|/(2\sqrt{\pi}), L=\emptyset$
\Ensure
$\mathbf{Rx}\in\Lambda$
\State invoke \textbf{Function 1} with $i=n$ to decode layer by layer
\State add all the candidates $\widehat{\mathbf{x}}$'s generated by \textbf{Function 1} to $L$
\State output $\widehat{\mathbf{x}}=\underset{\mathbf{x}\in L}{\operatorname{arg~min}} \, \|\mathbf{y}-\mathbf{R}\mathbf{x}\|$ as the decoding solution
\end{algorithmic}
\end{algorithm}

\floatname{algorithm}{Function}

\newcounter{TempEqCnt}                         
\setcounter{TempEqCnt}{\value{algorithm}} 
\setcounter{algorithm}{0}

\begin{algorithm}[t]
\caption{Searching at layer $i$ given $[\widehat{x}_n,\ldots,\widehat{x}_{i+1}]$}
\begin{algorithmic}[1]
\State compute $\widetilde{x}_i$ according to (\ref{eqn:noise effection in determination})
\State compute $f(\widehat{x}^j_i)$ by (\ref{fn1})
\State compute $K(\widehat{x}^j_i)$ according to (\ref{prunethresholdla})
\For {each specific integer candidate $\widehat{x}^j_i$}
\If {$K(\widehat{x}^j_i)<1$}
\State \hspace{-2em}prune $\widehat{x}^j_i$ from the tree searching
\Else
\State {\hspace{-2em}save $\widehat{x}^j_i$} to form the decoding result $[\widehat{x}_n,\ldots,\widehat{x}_{i+1}, \widehat{x}^j_i]$
\If {$i=1$}
\State \hspace{-2em}output the candidate $\widehat{\mathbf{x}}$
\Else
\State \hspace{-2em}invoke \textbf{Function 1} to search the next layer $i-1$
\EndIf
\EndIf
\EndFor
\end{algorithmic}
\end{algorithm}


Consequently, denoting $d(\mathbf{\Lambda},\mathbf{y})$ as the Euclidean distance between the given point $\mathbf{y}$ and lattice $\Lambda=\{\mathbf{Rx}: \mathbf{x}\in \mathbb{Z}^n\}$, consider to solve the ILS problem with sphere radius $D=d(\Lambda, \mathbf{y})$, the required initial searching size $K$ as well as the complexity $|S|$ can be derived in the following.

\begin{my2}
The required initial searching size $K$ of solving the ILS problem by ESD is $e^{2\pi d^2(\mathbf{\Lambda},\mathbf{y})/\min^2_{i}|r_{i,i}|}$, which corresponds to the complexity upper bounded by $|S|< n\cdot e^{2\pi d^2(\mathbf{\Lambda},\mathbf{y})/\min^2_{i}|r_{i,i}|}$.
\end{my2}

From Theorems 2 \& 3, the tractable decoding trade-off of ESD is obtained. Thanks to the usages of $K$ and $\sigma$, an insightful way is provided to reexamine the classic sphere decoding.
The operations of ESD are summarized in Algorithm 1. Note that Algorithm 1 entails a recursive decoding structure, but ESD also works in a non-recursive way, which enables a resource-efficient implementation on hardware like FPGA.

\section{Enhancement Mechanism I: Normalized Weighting}
The proposed ESD algorithm works based on the weighting $f(\widehat{x}^j_i)$ in (\ref{fn1}).
Now we show that ESD can be upgraded through the mechanism of normalized weighting, where a better decoding trade-off than that of Fincke-Pohst SD can be achieved.




\subsection{Normalized Weighting}
Specifically, we propose to replace the weighting $f(\widehat{x}^j_i)$ in ESD by a normalized one, defined as
\begin{equation}
p(\widehat{x}^j_i)\triangleq \frac{e^{-\frac{1}{2\sigma^2_i}\|\widehat{x}^j_i-\widetilde{x}_i\|^2}}{\sum_{\widehat{x}_i\in\mathbb{Z}}e^{-\frac{1}{2\sigma^2_i}\|\widehat{x}_i-\widetilde{x}_i\|^2}}=\frac{e^{-\frac{1}{2\sigma^2_i}\|\widehat{x}^j_i-\widetilde{x}_i\|^2}}{\rho_{\sigma_{i}, \widetilde{x}_{i}}(\mathbb{Z})},
\label{fn2}
\end{equation}
where
\begin{equation}
\rho_{\sigma_{i}, \widetilde{x}_{i}}(\mathbb{Z})=\sum_{\widehat{x}_i\in\mathbb{Z}}e^{-\frac{1}{2\sigma^2_i}\|\widehat{x}_i-\widetilde{x}_i\|^2}.
\end{equation}

Accordingly, the searching size $K(\widehat{x}^j_i)$ in (\ref{prunethresholdla}) is updated as
\begin{equation}
K(\widehat{x}^j_i)= K(\underline{\widehat{x}^j_{i}})\cdot p(\widehat{x}^j_i),
\label{prunethresholdl2}
\end{equation}
where the pruning threshold $K(\widehat{x}^j_i)\geq 1$ is retained as the same.
Clearly, normalized weighting $p(\cdot)$ offers a latent restriction, i.e.,
\begin{equation}
\sum_{\widehat{x}^j_i\in\mathbb{Z}} p(\widehat{x}^j_i)=1.
\end{equation}
Intuitively, by normalization --- the searching size $K(\underline{\widehat{x}^j_i})$ of a parent node can be viewed as to be reasonably allocated to its children nodes by (\ref{prunethresholdl2}), rather than diminished with $f(\cdot)$ in an exponential way.
Such a change is helpful to the searching in ESD by well retaining the searching size at each layer, so that more candidate nodes could be visited during the searching process.


\begin{my2}
With the normalized weighting $p(\cdot)$, the sphere radius achieved by ESD becomes
\end{my2}
\vspace{-1em}
\begin{equation}
D=\sigma\sqrt{2\ln \frac{K}{\prod^n_{i=1}\rho_{\sigma_{n-i+1}, \widetilde{x}_{n-i+1}}(\mathbb{Z})}}.
\label{mp1}
\end{equation}
\begin{proof}
The pruning threshold $K(\widehat{x}^j_i)\geq1$ with the normalized weighting $p(\cdot)$ can be expressed as
\begin{equation}
p(\widehat{x}^j_i)\geq\frac{1}{K(\underline{\widehat{x}^j_i})}=\frac{1}{K\cdot p(\widehat{x}^j_{i+1})\cdots p(\widehat{x}^j_{n})}.
\label{prunethreshold22}
\end{equation}

From (\ref{prunethreshold22}), for any candidate vector $\widehat{\mathbf{x}}$ being obtained by ESD, its normalized weighting $p(\widehat{x}_1)$ of element $\widehat{x}_1$ at the layer $i=1$ must satisfy
\begin{equation}
p(\widehat{x}_1)\geq\frac{1}{K\cdot p(\widehat{x}_{2})\cdots p(\widehat{x}_{n})},
\label{b12b1}
\end{equation}
which results in the following lower bound
{\allowdisplaybreaks\begin{flalign}
\prod^n_{i=1}p(\widehat{x}_{n-i+1})&=\prod^n_{i=1}\frac{e^{-\frac{1}{2\sigma^2_{n-i+1}}\|\widehat{x}_{n-i+1}-\widetilde{x}_{n-i+1}\|^2}}{\sum_{\widehat{x}_{n-i+1}\in\mathbb{Z}}e^{-\frac{1}{2\sigma^2_{n-i+1}}\|\widehat{x}_{n-i+1}-\widetilde{x}_{n-i+1}\|^2}}\notag\\
&=\frac{e^{-\frac{1}{2\sigma^2}\|\mathbf{R}\mathbf{x}-\mathbf{y}\|^2}}{\prod^n_{i=1}\rho_{\sigma_{n-i+1}, \widetilde{x}_{n-i+1}}(\mathbb{Z})}\notag\\
&\geq \frac{1}{K}.
\end{flalign}}Then, by simple transformation, the above lower bound corresponds to obtaining the candidate vectors $\widehat{\mathbf{x}}$'s within the sphere radius
\begin{equation}
\|\mathbf{R}\mathbf{x}-\mathbf{y}\|\leq \sigma\sqrt{2\ln \frac{K}{\prod^n_{i=1}\rho_{\sigma_{n-i+1}, \widetilde{x}_{n-i+1}}(\mathbb{Z})}},
\label{b12a2}
\end{equation}
completing the proof.
\end{proof}

Next, we show that the normalized weighting $p(\cdot)$ turns out to be a better choice than the original weighting $f(\cdot)$ in terms of sphere radius.

\begin{my7}
For $\sigma\leq\min_{i}|r_{i,i}|/(2\sqrt{\pi})$, ESD with normalized weighting $p(\cdot)$ achieves a larger sphere radius than that with weighting $f(\cdot)$ due to
\end{my7}
\vspace{-1em}
\begin{equation}
\prod^n_{i=1}\rho_{\sigma_{n-i+1}, \widetilde{x}_{n-i+1}}(\mathbb{Z})\leq1.
\end{equation}
\begin{proof}
Similar to the derivation in (\ref{pj1a1}), it is straightforward to verify that
{\allowdisplaybreaks\begin{flalign}
\rho_{\sigma_i, \widetilde{x}_i}(\mathbb{Z})&\leq \rho_{\sigma_i}(\mathbb{Z})\notag\\
&=\sum_{\widehat{x}_i\in\mathbb{Z}}e^{-\frac{1}{2\sigma^2_i}\|\widehat{x}_i\|^2}\label{9u2}\\
&= \vartheta_3(|r_{i,i}|^2/2\pi\sigma^2)\notag\\
&\approx1,
\end{flalign}completing the proof.}
\end{proof}

Note that the equality in (\ref{9u2}) only holds when $\widetilde{x}_i\in\mathbb{Z}$. Consider the product from $i=n$ to $i=1$, the case of $\prod^n_{i=1}\rho_{\sigma_{n-i+1}, \widetilde{x}_{n-i+1}}(\mathbb{Z})=\prod^n_{i=1}\rho_{\sigma_{n-i+1}}(\mathbb{Z})$ tends to rarely happen, implying the remarkable superiority of $p(\cdot)$ over $f(\cdot)$.
Given the target sphere radius $D=d(\Lambda, \mathbf{y})$, it is straightforward to see that ML decoding performance can be achieved if
\begin{equation}
K\geq \left(\prod^n_{i=1}\rho_{\sigma_{n-i+1}, \widetilde{x}_{n-i+1}}(\mathbb{Z})\right)\cdot e^{2\pi d^2(\mathbf{\Lambda},\mathbf{y})/\min^2_{i}|r_{i,i}|}.
\end{equation}
Note that $K$ is much smaller than the required $K$ with $f(\cdot)$.

\subsection{Complexity Analysis}
Next, we study the complexity of ESD with normalized weighting $p(\cdot)$.



\begin{my1}
In ESD with normalized weighting $p(\cdot)$, for each parent candidate node $\underline{\widehat{x}^j_{i}}$ with $K(\underline{\widehat{x}^j_i})\geq1$,
the number of its saved children candidate nodes at decoding layer $i$ satisfies
\end{my1}
\vspace{-1em}
\begin{equation}
K_{\text{save}}\leq K(\underline{\widehat{x}^j_i})
\label{pj1a}
\end{equation}
\emph{if} $\sigma=\min_{i}|r_{i,i}|/(2\sqrt{\pi})$.
\begin{proof}
We start the proof by considering the cases of $1\leq K(\underline{\widehat{x}^j_i})<2$ and $K(\underline{\widehat{x}^j_i})\geq2$ respectively.

On one hand, based on the pruning threshold in (\ref{prunethreshold22}), candidate nodes with $1\leq K(\underline{\widehat{x}^j_i})<2$ will be saved if
\begin{equation}
p(\widehat{x}^j_i)\geq\frac{1}{K(\underline{\widehat{x}^j_i})}>\frac{1}{2}.
\label{owq1}
\end{equation}
Clearly, because of $\sum_jp(\widehat{x}^j_i)=1$, there is at most one integer candidate node satisfying (\ref{owq1}), implying
\begin{equation}
K_{\text{save}}\leq1\leq K(\underline{\widehat{x}^j_i})
\end{equation}
no matter what $\sigma>0$ is.

On the other hand, when $K(\underline{\widehat{x}^j_i})\geq2$, according to the pruning threshold in (\ref{prunethreshold22}), the condition shown in (\ref{pj1a}) holds if and only if the $\lfloor K(\underline{\widehat{x}^j_i})+1\rfloor$th closest integer candidate node to $\widetilde{x}_i$ is definitely pruned, that is
\begin{equation}
K(\underline{\widehat{x}^j_i})p(\widehat{x}^{\lfloor K(\underline{\widehat{x}^j_i})+1 \rfloor}_i)< 1.
\label{wo1a1}
\end{equation}

%
%

Then, from (\ref{wo31}), (\ref{wo1a1}) can be achieved if
\begin{equation}
K(\underline{\widehat{x}^j_i})\cdot e^{-\frac{1}{8\sigma_i^2}(\lfloor K(\underline{\widehat{x}^j_i})+1\rfloor-1)^2}<\rho_{\sigma_{i}, \widetilde{x}_{i}}(\mathbb{Z}).
\label{wo4a}
\end{equation}
Moreover, according to the following relationship \cite{RegevNP}
\begin{equation}
\rho_{\sigma_{i}, \widetilde{x}_{i}}(\mathbb{Z})\geq e^{-\frac{d^2(\mathbb{Z}, \widetilde{x}_i)}{2\sigma^2_i}}\cdot\rho_{\sigma_{i}}(\mathbb{Z})
\label{xpq1c}
\end{equation}
with $d(\mathbb{Z}, \widetilde{x}_i)$ denoting the Euclidean distance between $\widetilde{x}_i$ and its closest integer over $\mathbb{Z}$, (\ref{wo4a}) holds if
\begin{equation}
K(\underline{\widehat{x}^j_i})\cdot e^{-\frac{1}{8\sigma_i^2}(\lfloor K(\underline{\widehat{x}^j_i})+1\rfloor-1)^2}\hspace{-.5em}<\hspace{-.1em}e^{-\frac{d^2(\mathbb{Z}, \widetilde{x}_i)}{2\sigma^2_i}}\hspace{-.2em}\cdot\rho_{\sigma_{i}}(\mathbb{Z})
\label{wo4b}
\end{equation}
is fulfilled. Because of $0\leq d(\mathbb{Z}, \widetilde{x}_i)\leq1/2$, (\ref{wo4b}) becomes
\begin{equation}
\sigma^2<\frac{(\lfloor K(\underline{\widehat{x}^j_i})+1\rfloor-1)^2-1}{8\ln (K(\underline{\widehat{x}^j_i})/\rho_{\sigma_{i}}(\mathbb{Z}))}\cdot\|\widehat{\mathbf{b}}_i\|^2.
\label{wo5a}
\end{equation}


Consequently, it is clear to verify that (\ref{wo5a}) is satisfied when $\sigma=\min_{i}|r_{i,i}|/(2\sqrt{\pi})$ (i.e., $\rho_{\sigma_{i}}(\mathbb{Z})=\sum_{\widehat{x}_i\in\mathbb{Z}}e^{-\frac{1}{2\sigma^2_i}\|\widehat{x}_i\|^2}\approx1$). This completes the proof.
\end{proof}

\begin{my1}
In ESD with normalized weighting $p(\cdot)$, for each parent candidate node $\underline{\widehat{x}^j_{i}}$ with $K(\underline{\widehat{x}^j_i})\geq1$,
the summation of searching sizes of its saved children candidate nodes at decoding layer $i$ is decreasing
\end{my1}
\vspace{-1em}
\begin{equation}
\sum_jK(\widehat{x}^j_i)< K(\underline{\widehat{x}^j_i})
\label{pj1a1}
\end{equation}
if $\sigma\leq\min_{i}|r_{i,i}|/(2\sqrt{\pi})$.
\begin{proof}
By (\ref{prunethresholdl2}), for each parent candidate node $\underline{\widehat{x}^j_{i}}$ with $K(\underline{\widehat{x}^j_i})\geq1$, the summation of searching sizes of its saved children candidate nodes follows
\begin{equation}
\sum_j\hspace{-.2em}K(\widehat{x}^j_i\hspace{-.1em})\hspace{-.3em}=\hspace{-.3em}K(\underline{\widehat{x}^j_i}\hspace{-.1em})\cdot\hspace{-.1em}\sum_j\hspace{-.2em}p(\widehat{x}^j_i\hspace{-.1em})\hspace{-.2em}<\hspace{-.3em}K(\underline{\widehat{x}^j_i}\hspace{-.1em})\cdot\hspace{-.2em}\sum_{\widehat{x}^j_i\in\mathbb{Z}} \hspace{-.3em}p(\widehat{x}^j_i\hspace{-.1em})\hspace{-.2em}=\hspace{-.2em}K(\underline{\widehat{x}^j_i}\hspace{-.1em}).
\end{equation}
Here, the inequality holds since partially searching sizes would be discarded as their children nodes fail to satisfy the pruning threshold.
\end{proof}

From Lemma 4 \& 5, the number of visited nodes of ESD with normalized weighting $p(\cdot)$ can be derived as follows.
\begin{my2}
In ESD with normalized weighting $p(\cdot)$, let $\sigma=\min_{i}|r_{i,i}|/(2\sqrt{\pi})$, the number of visited nodes is upper bounded by
\end{my2}
\vspace{-2em}
\begin{equation}
|S|< nK,
\label{hp6a}
\end{equation}
\emph{and the number of collected candidate vectors is upper bounded by}
\begin{equation}
|L|< K.
\label{hp6c}
\end{equation}
\begin{proof}
According to (\ref{pj1a}), the number of saved candidate nodes at each layer is upper bounded by the summation of pruning sizes at the previous layer, namely,
\begin{equation}
K^{\text{layer}\ i}_{\text{save}}=\sum K_{\text{save}}\leq \sum K(\underline{\widehat{x}^j_i})=K^{\text{layer}\ i+1}_{\text{search\ size}}.
\label{58a}
\end{equation}


Then, by (\ref{pj1a1}), it is easy to confirm the summation of searching sizes at each layer is decreasing from layer $n$ to 1, i.e.,
\begin{equation}
K^{\text{layer}\ 1}_{\text{search\ size}}<\ldots< K^{\text{layer}\ n}_{\text{search\ size}}< K^{\text{layer}\ n+1}_{\text{search\ size}}=K.
\label{pj1pa1}
\end{equation}



Therefore, the number of visited nodes is upper bounded by
\begin{equation}
|S|=\sum_iK^{\text{layer}\ i}_{\text{save}}\leq \sum_iK^{\text{layer}\ i+1}_{\text{search\ size}}< nK.
\end{equation}
Moreover, since the number of collected searching candidates $|L|$ accounts for $K^{\text{layer}\ 1}_{\text{save}}$, it is upper bounded by
\begin{equation}
|L|< K.
\end{equation}
\end{proof}

Based on Corollary 3 and Theorems 5, when $\sigma=\min_{i}|r_{i,i}|/(2\sqrt{\pi})$, ESD with normalized weighting $p(\cdot)$ achieves a larger sphere radius than the one with weighting function $f(\cdot)$ (i.e., Fincke-Pohst SD) under the same complexity upper bound, thus leading to a better decoding trade-off between performance and complexity.

Another point should be emphasized here is that normalized weighting is well suited to the cases of finite state space, i.e., $\mathbf{x}\in\mathcal{X}^n$, $\mathcal{X}\in\mathbb{Z}$.
For the limited state space of $\mathcal{X}$, the searching size $K(\widehat{x}^j_i)$ based on $f(\cdot)$ may vanish rapidly if $\widetilde{x}_i$ from (\ref{eqn:noise effection in determination}) locates outside of $\mathcal{X}$, thus terminating the searching at the very early stages.
On the contrary, according to the normalized weighting $p(\cdot)$, such a risk could be effectively alleviated as the searching size $K(\widehat{x}^j_i)$ could be well retained. In particular, if $\widetilde{x}_i$ is far away from $\mathcal{X}$, the closest integer candidate node $\widehat{x}^j_i\in\mathcal{X}$ to $\widetilde{x}_i$ will be saved with the overwhelming normalized weighting $p(\widehat{x}^j_i)$. In this way, most of the searching size will be retained for the subsequent searching rather than be vanished, which results in a better decoding performance.
Further complexity reduction about the choice of selected nodes in practice is considered in Section VI.



\section{Enhancement Mechanism II: Candidate Protection}
From (\ref{pj1pa1}), since the summation of searching sizes is decreasing layer by layer, there is a latent problem in the proposed ESD algorithm: it only works well when the initial searching size $K$ is large enough.
Given a small size $K$, the searching still works but it will terminate at the very early layers because all the possible candidate nodes are pruned by the small searching size $K(\widehat{x}^j_i)$.
This is similar to Fincke-Pohst SD, where no decoding solution will be outputted by a small sphere radius $D$.
In this case, although considerable complexity cost has been consumed, no eligible candidate vector $\widehat{\mathbf{x}}$ will be returned, rendering the searching meaningless.
This actually raises a critical question to ESD: how to fully exploit the decoding potential with a small or moderate $K$?
Next, we try to answer this question via another enhancement mechanism designed for ESD, and we refer to it as \emph{candidate protection}.

\subsection{Candidate Protection}
In essence, as for candidate nodes with small searching size $K(\widehat{x}^j_i)$, the mechanism of candidate protection tries to rescue the most valuable candidate vector along that searching branch, and
the searching solution consists of the closest candidate nodes $\widehat{x}^1_{i's}$ in the rest of layers normally turns out to be perspective in statistics.


Specifically, as for candidate node $\widehat{x}^j_i$ with small searching size
\begin{equation}
2> K(\widehat{x}^j_i)\geq 1,
\end{equation}
candidate protection is activated to obtain the closest integer nodes $\widehat{x}^1_{i-1}, \ldots, \widehat{x}^1_1$ in the rest of searching layers, which directly yields a candidate vector $\widehat{\mathbf{x}}$:
\begin{equation}
\widehat{\mathbf{x}}=[\ \overbrace{\underbrace{\widehat{x}^1_1\ ,\ \ldots ,\ \widehat{x}^1_{i-1}}_{\text{candidate protection}}\ , \hspace{-.8em}\underbrace{\widehat{x}^{j}_i}_{2> K(\cdot)\geq 1}\hspace{-1em},\ \underbrace{\widehat{x}^{j}_{i+1},\ \ldots,\ \widehat{x}^{j}_n}_{K(\cdot)\geq2}}^{\longleftarrow\text{decoding order}}\ ]^T.
\label{pwu1}
\end{equation}
For a better understanding, Fig. 3 illustrates the operations of candidate protection.
\begin{figure}[h]
\begin{center}
\hspace{-1em}\includegraphics[width=2.6in]{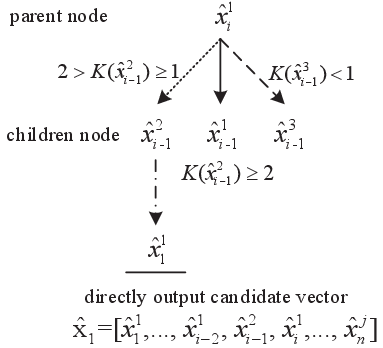}
\end{center}
\vspace{-1em}
  \caption{The illustration of candidate protection, where node $\widehat{x}^2_{i-1}$ invokes candidate protection to directly output a candidate vector $\widehat{\mathbf{x}}$ to set $L$.}
  \label{simulation x1}
\end{figure}

We point out that the pruning threshold $K(\widehat{x}^j_i)\geq 1$ is smoothly compatible with candidate protection as the latter tries to activate a few candidate nodes discarded by the former.
Intuitively, the proposed candidate protection extends the initial searching size from $K>1$ to $K\geq1$, and it is easy to verify that the decoding performance of Babai's nearest plane algorithm will be achieved when $K=1$.
More specifically, candidate protection can be simply carried out through Babai's nearest plane algorithm since $[\widehat{x}^1_1,...,\widehat{x}^1_n]^T$ is just the decoding result of it.



\begin{my2}
For ESD with normalized weighting and candidate protection, flexible decoding performance can be achieved from Babai's nearest plane algorithm (i.e., $K=1$) and ML decoding (i.e., $K\geq \left(\prod^n_{i=1}\rho_{\sigma_{n-i+1}, \widetilde{x}_{n-i+1}}(\mathbb{Z})\right)\cdot e^{2\pi d^2(\mathbf{\Lambda},\mathbf{y})/\min^2_{i}|r_{i,i}|}$).
\end{my2}

To summarize, at each searching layer, ESD with normalized weighting and candidate protection operates in the following two steps:
\begin{itemize}
  \item \emph{Calculate the searching size $K(\widehat{x}_i^j)$ by (\ref{prunethresholdl2})}.
  \item \emph{Obtain candidate nodes $\widehat{x}_i^j$ by (\ref{prunethreshold1b}). If $2> K(\widehat{x}^j_i)\geq 1$, invoke Babai's nearest plane algorithm to directly return a decoding candidate vector $\widehat{\mathbf{x}}$}.
\end{itemize}
An illustration of the proposed ESD algorithm is presented in Fig. 4 with more details.
In addition, we claim that candidate protection can also be applied to ESD with weighting $f(\cdot)$ to yield suboptimal decoding solutions.

\begin{figure*}[t]
\includegraphics[width=7.2in,height=3.3in]{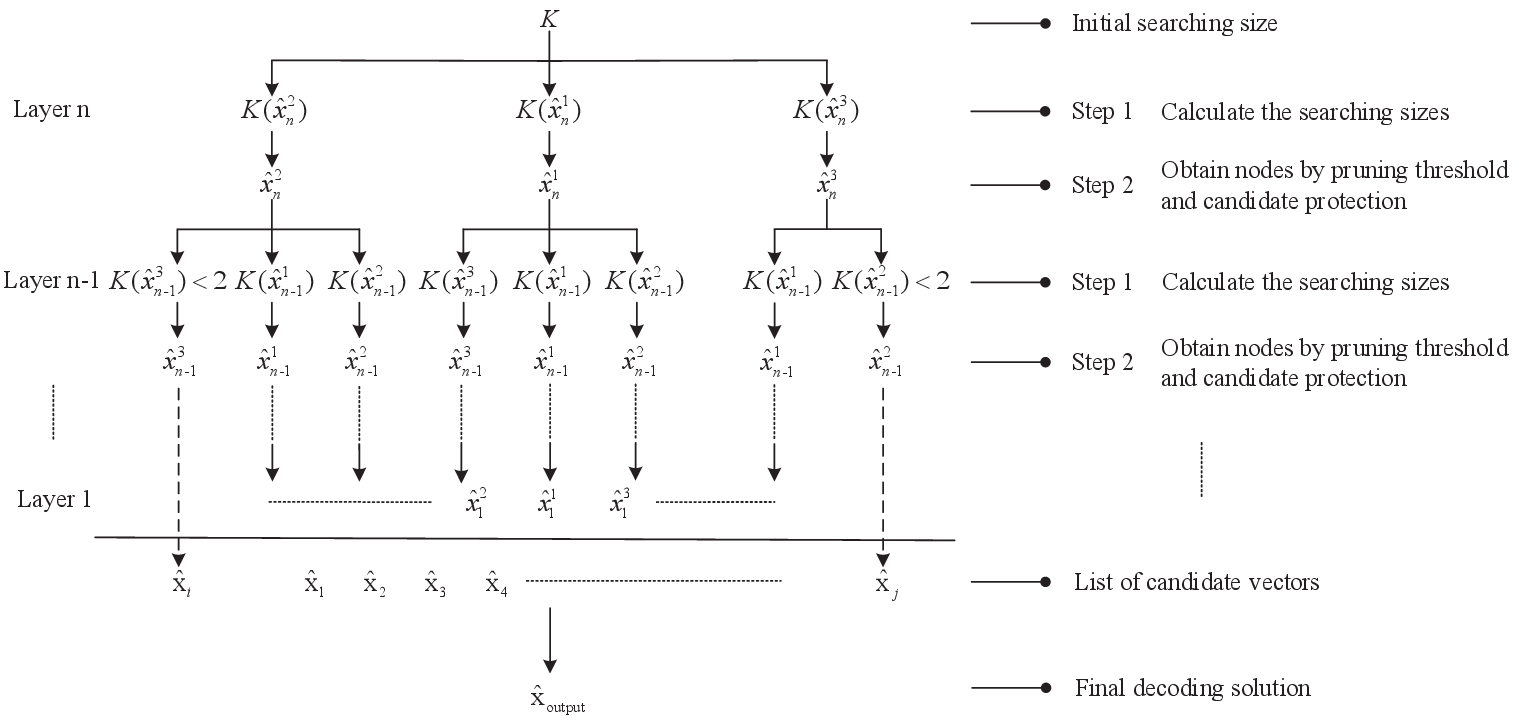}
\caption{Illustration of the proposed ESD algorithm with normalized weighting and candidate protection, where $K(\widehat{x}^j_i)\geq1$. The dashed lines stemmed from $K(\widehat{x}^j_i)<2$ denote the closest candidate nodes $\widehat{x}^1_{i-1}, \ldots, \widehat{x}^1_1$ in the rest of layers, which are retained to directly yield a decoding candidate vector $\widehat{\mathbf{x}}$.}
\label{IDD figure}
\end{figure*}

\subsection{Complexity Analysis}
Interestingly, even with normalized weighting and candidate protection, the complexity $|S|$ as well as the number of collected candidate vectors $|L|$ in ESD still maintains the same upper bound as before.
\begin{my2}
Given the initial searching size $K\geq1$, the number of candidate vectors collected by ESD with normalized weighting and candidate protection is upper bounded by
\end{my2}
\vspace{-1.5em}
\begin{equation}
|L|< K
\label{pj1pala}
\end{equation}
\emph{with the bounded number of visited nodes}
\begin{equation}
|S|< nK
\label{pj1palb}
\end{equation}
\vspace{-.5em}
\emph{for} $\sigma=\min_{i}|r_{i,i}|/(2\sqrt{\pi})$.
\vspace{1em}
\begin{proof}
Theoretically, the collected candidate vectors $\widehat{\mathbf{x}}$'s come from pruning threshold and candidate protection respectively.
For notational simplicity, here we represent the searching size $K(x^j_i)$ with two different ways: $2>K(x^{\text{protection}}_i)\geq1$ and $K(x^{\text{pruning}}_{i})\geq2$.

Specifically, the summation of the searching sizes at each layer is decreasing, which can be expressed as
{\allowdisplaybreaks\begin{flalign}
K&=K(\underline{x^j_n}) \notag\\
&>\sum K(x^{\text{protection}}_n)+\sum K(x^{\text{pruning}}_n) \notag\\
&>\sum K(x^{\text{protection}}_n)+\sum K(x^{\text{protection}}_{n-1})+\sum K(x^{\text{pruning}}_{n-1}) \notag\\
&>\cdots\notag\\
&>\sum_{i=2}^n\left[\sum K(x^{\text{protection}}_i)\right]+\sum K(x^{\text{pruning}}_{2}).
\label{59a}
\end{flalign}}

Based on candidate protection, only one decoding candidate vector will be saved for each $K(x^{\text{protection}}_i)$, $2\leq i\leq n$, which means the number of collected candidate vectors generated by candidate protection from searching layer $n$ to $2$ is bounded by
\begin{equation}
|L_{\text{protection}}|\leq\sum_{i=2}^n\left[\sum K(x^{\text{protection}}_i)\right].
\label{k1a}
\end{equation}

Besides, the number of candidate vectors survived from the pruning threshold corresponds to the number of saved candidate nodes at layer $i=1$, i.e., $K^{\text{layer}\ 1}_{\text{save}}$, which is upper bounded by
\begin{equation}
|L_{\text{pruning}}|=K^{\text{layer}\ 1}_{\text{save}}\leq \sum K(x^{\text{pruning}}_{2})
\label{k1b}
\end{equation}
according to (\ref{58a}).
Therefore, based on (\ref{59a}), (\ref{k1a}) and (\ref{k1b}), there is
\begin{equation}
|L|=|L_{\text{pruning}}|+|L_{\text{protection}}|<K.
\end{equation}
Consequently, as all the visited nodes are taken into account to generate $|L|$ decoding candidate vectors, the number of visited nodes is bounded as
\begin{equation}
|S|< n|L|< nK,
\end{equation}
completing the proof.
\end{proof}



\renewcommand{\arraystretch}{1.8}
\begin{table*}[t]
\begin{center}
\caption{Performance and Complexity of various decoding schemes.}
\label{tab:TCQvsPL}
\begin{tabular}{|c||c||c|}\hline
 & Decoding Radius & Number of Visited Nodes \\\hline
$\text{Klein Sampling}$ \cite{Klein}& $\sqrt{\log_n(Ke^{-2})}\cdot\min_{i}|r_{i,i}|$  & $|S|=nK$  \\\hline
$\text{Randomized Sampling}$ \cite{CongRandom}& $\sqrt{\log_\varrho(Ke^{-2n/\varrho})}\cdot\min_{i}|r_{i,i}|$,\ $\varrho>1$  & $|S|=nK$ \\\hline
$\text{IMHK Sampling}$ \cite{ZhengWangTIT17}& $\sqrt{(\ln\frac{K}{\log(1/\epsilon)\cdot\prod^n_{i=1}\hspace{-.2em}\rho_{\sigma_{n\hspace{-.1em}-\hspace{-.1em}i\hspace{-.1em}+\hspace{-.1em}1}, \widetilde{x}_{n\hspace{-.1em}-\hspace{-.1em}i\hspace{-.1em}+\hspace{-.1em}1}}(\mathbb{Z})})/(2\pi)}\cdot\min_{i}|r_{i,i}|$ & $|S|=nK$\\\hline
$\text{ESD with}\ f(\cdot)$ (Cor. 1) & $\sqrt{(\ln K)/(2\pi)}\cdot\min_{i}|r_{i,i}|$  & $|S|<nK$ \\\hline
$\text{ESD with}\ p(\cdot)$ (Thm. 4) & $\sqrt{(\ln\frac{K}{\prod^n_{i=1}\hspace{-.2em}\rho_{\sigma_{n\hspace{-.1em}-\hspace{-.1em}i\hspace{-.1em}+\hspace{-.1em}1}, \widetilde{x}_{n\hspace{-.1em}-\hspace{-.1em}i\hspace{-.1em}+\hspace{-.1em}1}}(\mathbb{Z})})/(2\pi)}\cdot\min_{i}|r_{i,i}|$ & $|S|<nK$ \\\hline
$\text{ESD based on LGD}$ & $\sqrt{(\ln\frac{K}{\rho_{\sigma,\mathbf{y}}(\Lambda)})/(2\pi)}\cdot\min_{i}|r_{i,i}|$ & N/A\\\hline
\end{tabular}
\end{center}
\end{table*}

\section{Performance Optimization and Complexity Reduction}
In this section, further performance optimization and complexity reduction with respect to ESD are investigated. Meanwhile, the relationship between ESD and lattice Gaussian sampling decoding is also revealed.

\subsection{The Perspective of ESD over Lattice Gaussian Distribution}

From (\ref{b12a2}), the paradigm of ESD with $p(\cdot)$ can be described as
\begin{equation}
\frac{e^{-\frac{1}{2\sigma^2}}\|\mathbf{Rx}-\mathbf{y}\|^2}{\prod^n_{i=1}\rho_{\sigma_{n-i+1}, \widetilde{x}_{n-i+1}}(\mathbb{Z})}\geq \frac{1}{K},
\label{xnb1}
\end{equation}
where the LHS of (\ref{xnb1}) can be viewed as a Gaussian-like distribution regarding to $\mathbf{x}$, i.e.,
\begin{equation}
\mathcal{G}(\mathbf{x})=\frac{e^{-\frac{1}{2\sigma^2}}\|\mathbf{Rx}-\mathbf{y}\|^2}{\prod^n_{i=1}\rho_{\sigma_{n-i+1}, \widetilde{x}_{n-i+1}}(\mathbb{Z})}.
\label{xnc2}
\end{equation}
Intuitively, we can interpret ESD as enumerating all the possible candidate vectors $\widehat{\mathbf{x}}$'s with probabilities above a certain level (i.e., $\mathcal{G}(\mathbf{x})\geq1/K$).

On the other hand, different from the enumerating in ESD, sampling from $\mathcal{G}(\mathbf{x})$ also provides a feasible way to solve the ILS problem.
In particular, by multiple independent samplings over $\mathcal{G}(\mathbf{x})$, the optimal solution $\widehat{\mathbf{x}}_{\text{ML}}$ would most likely be obtained due to its relatively large sampling probability.
Fortunately, it has been shown in \cite{Klein} that $\mathcal{G}(\mathbf{x})$ can be sampled by Klein's sampling algorithm, and the randomized sampling decoding scheme is further proposed and investigated in \cite{CongRandom,DerandomizedJ}.
Nevertheless, inevitable performance loss does exist due to the distortion of the Gaussian-like distribution, as the target optimal solution $\widehat{\mathbf{x}}_{\text{ML}}$ in MIMO detection is not guaranteed to have the largest sampling probability in $\mathcal{G}(\cdot)$.

Recently, the concept of lattice Gaussian distribution was proposed, i.e.,
\begin{equation}
\mathcal{D}_{\Lambda,\sigma,\mathbf{y}}(\mathbf{x})\hspace{-.1em}=\hspace{-.1em}\frac{e^{-\frac{1}{2\sigma^2}\parallel \mathbf{Rx}-\mathbf{y} \parallel^2}}{\sum_{\mathbf{x} \in \mathbb{Z}^n}e^{-\frac{1}{2\sigma^2}\parallel \mathbf{Rx}-\mathbf{y} \parallel^2}}\hspace{-.1em}=\hspace{-.1em}\frac{e^{-\frac{1}{2\sigma^2}\parallel \mathbf{Rx}-\mathbf{y} \parallel^2}}{\rho_{\sigma,\mathbf{y}}(\Lambda)},
\label{lattice gaussian distribution}
\end{equation}
which has been a central role in various research fields\footnote{The parameter $\sigma>0$ in lattice Gaussian distribution $\mathcal{D}_{\Lambda,\sigma,\mathbf{y}}(\mathbf{x})$ is known as standard deviation.} \cite{Banaszczyk,Forney_89,LiuLing2,LB_13,LiuLing1,LLBS_12,7058433,ProbabilisticShapingOptical1}.
As for solving the ILS problem, the exact Gaussian distribution $\mathcal{D}_{\Lambda,\sigma,\mathbf{y}}(\mathbf{x})$ turns out to be a better choice than the Gaussian-like distribution $\mathcal{G}(\mathbf{x})$
because the optimal decoding solution with the smallest Euclidean distance naturally entails the largest probability to be sampled, namely,
\begin{equation}
\widehat{\mathbf{x}}_{\text{ML}}=\underset{\mathbf{x}\in \mathbb{Z}^{n}}{\operatorname{arg~min}} \, \|\mathbf{R}\mathbf{x}-\mathbf{y}\|^2=\underset{\mathbf{x}\in \mathbb{Z}^{n}}{\operatorname{arg~max}}\ \mathcal{D}_{\Lambda,\sigma,\mathbf{y}}(\mathbf{x}).
\end{equation}
Such an equivalence can be found in Fig. 5, where sampling or enumeration over $\mathcal{D}_{\Lambda,\sigma,\mathbf{y}}(\mathbf{x})$ can be carried out to obtain the target $\widehat{\mathbf{x}}_{\text{ML}}$.
From this perspective, sampling or enumeration over $\mathcal{G}(\mathbf{x})$ can be viewed as an approximation of it.
However, in sharp contrast to the continuous Gaussian density, it is by no means trivial even to sample from a low-dimensional discrete Gaussian distribution.
To achieve the sampling from $\mathcal{D}_{\Lambda,\sigma,\mathbf{y}}(\mathbf{x})$, Markov chain Monte Carlo (MCMC) methods have been introduced, and the independent Metropolis-Hastings-Klein (IMHK) sampling algorithm was proposed to perform the sampling through a sophisticated Markov chain \cite{ZhengWangTIT15,ZhengWangTIT17}.

\begin{figure}[h]
\begin{center}
\hspace{-1em}\includegraphics[width=3.6in]{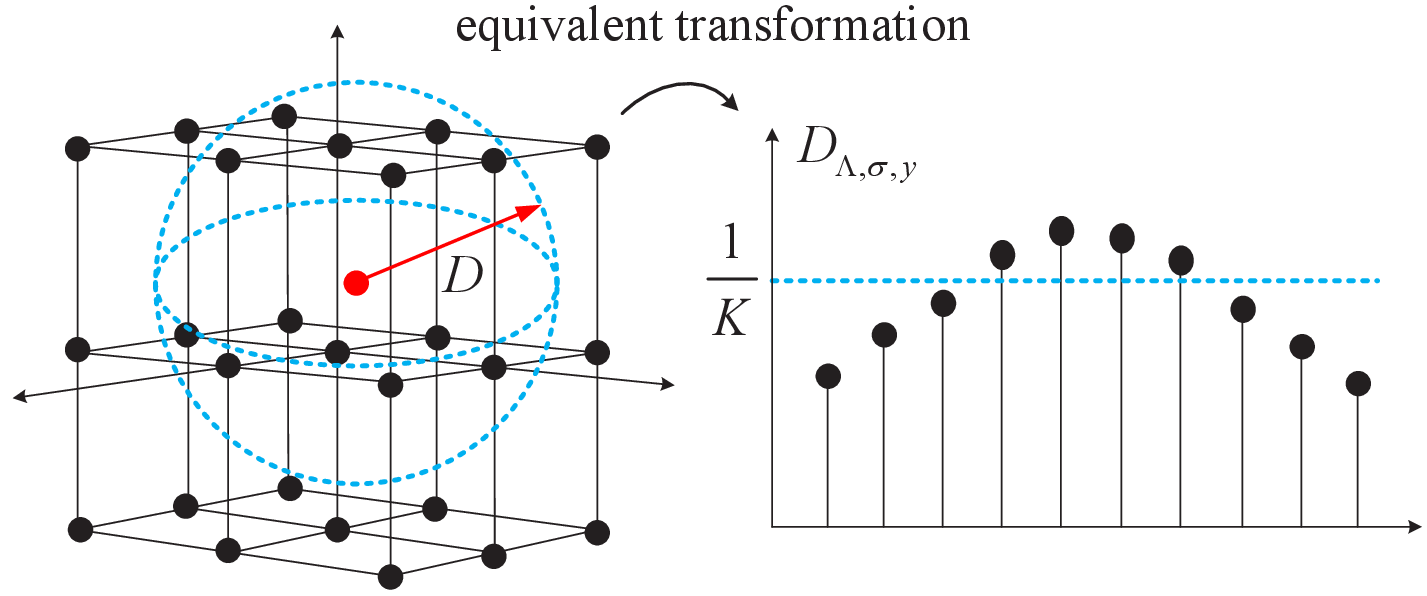}
\end{center}
\vspace{-1em}
  \caption{The equivalent transformation of solving the 3-dimensional ILS problem from Euclidean distance to probability in lattice Gaussian distribution.}
  \label{simulation x}
\end{figure}

It is clear that sampling decoding over lattice Gaussian distribution $\mathcal{D}_{\Lambda,\sigma,\mathbf{y}}(\mathbf{x})$ also entails a flexible decoding trade-off determined by the sampling number $K$\footnote{For straightforward comparison, here we also use $K$ to denote the number of sampling in sampling decoding.}.
However, compared to the deterministic enumeration in ESD, sampling decoding suffers from inevitable performance loss and complexity waste due to the inherent randomness during the sampling.
To make it clear, with $\sigma=\min_{i}|r_{i,i}|/(2\sqrt{\pi})$, the comparison over decoding radius and the number of visited nodes are summarized in Table I.
Here, decoding radius is a concept from lattice decoding to evaluate the decoding performance \cite{CongProxity}. Typically, in ESD, decoding radius is the same as sphere radius. As for sampling decoding, ILS or CVP problem
will most likely be addressed if $d(\mathbf{\Lambda},\mathbf{y})$ is less than $\text{decoding radius}$ while the uncertainty mainly comes from the randomness during the sampling\footnote{More details about the decoding radius of sampling decoding can be found in \cite{CongRandom,ZhengWangTIT17}.}.
As can be seen clearly, ESD with $p(\cdot)$ outperforms sampling decoding schemes due to larger decoding radius and less complexity cost.
Note that the above decoding radii of ESD algorithms are only based on the survived candidate vectors collected by the pruning threshold, which means the real decoding performance could be better under the help of candidate protection.
On the other hand, the computational complexity of randomized sampling decoding is $O(Kn^2)$, which can serve as an upper bound for the proposed ESD algorithm.
Overall, we emphasize that remarkable decoding potential still does exist since ESD with $p(\cdot)$ actually performs the enumeration based on the Gaussian-like distribution $\mathcal{G}(\mathbf{x})$ rather than the Gaussian distribution $\mathcal{D}_{\Lambda,\sigma,\mathbf{y}}(\mathbf{x})$.

\begin{figure}[h]
\includegraphics[width=3.5in]{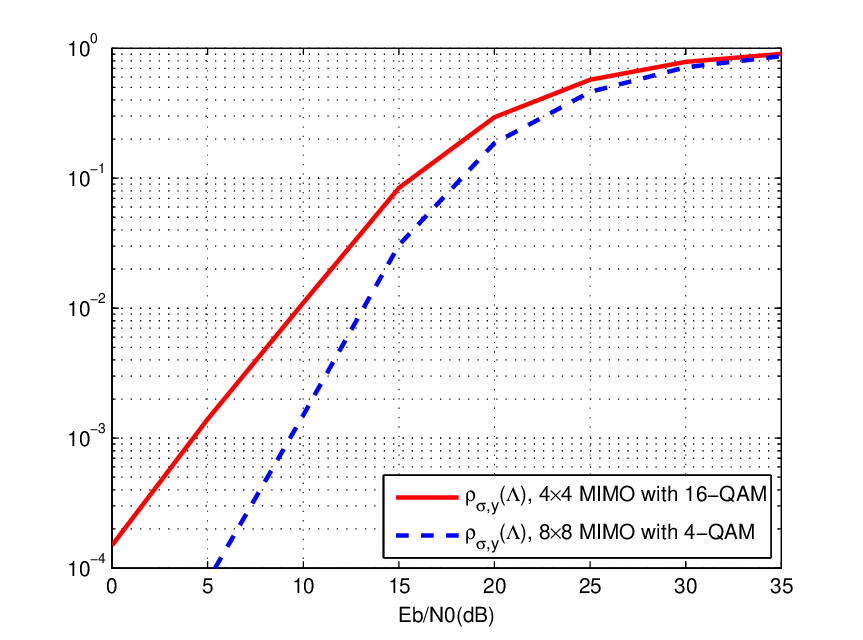}
\vspace{-1em}
  \caption{The Gaussian scalar $\rho_{\sigma,\mathbf{y}}(\Lambda)$ in various uncoded MIMO systems.}
  \label{simulation 2m}
\end{figure}


Unfortunately, the Gaussian scalar $\rho_{\sigma,\mathbf{y}}(\Lambda)=\sum_{\mathbf{x} \in \mathbb{Z}^n}e^{-\frac{1}{2\sigma^2}\parallel \mathbf{Rx}-\mathbf{y} \parallel^2}$ in $\mathcal{D}_{\Lambda,\sigma,\mathbf{y}}(\mathbf{x})$ is difficult to compute and factorize. In this condition, how to design the related searching algorithm by fully incorporating parameters $K$ and $\sigma$ turns out to be quite challenging. Otherwise, the enumeration regarding to $\mathcal{D}_{\Lambda,\sigma,\mathbf{y}}(\mathbf{x})$ can only be carried out by the conventional sphere decoding with sphere radius $D=\sqrt{(\ln\frac{K}{\rho_{\sigma,\mathbf{y}}(\Lambda)})/(2\pi)}\cdot\min_{i}|r_{i,i}|$, which fails to take advantages of the extra degrees of freedom from $K$ and $\sigma$.
Nevertheless, the sphere decoding over $\mathcal{D}_{\Lambda,\sigma,\mathbf{y}}(\mathbf{x})$ still provides a meaningful clue to the development of ESD for a better decoding trade-off.

To make it clear, the Gaussian scalar $\rho_{\sigma,\mathbf{y}}(\Lambda)$ with $\sigma=\min_{i}|r_{i,i}|/(2\sqrt{\pi})$ in MIMO scenarios is presented in Fig. 6 by Monte Carlo methods, where $\mathbf{x}\in\mathcal{X}^n$ belongs to a finite state space following QAM modulation.
Given Table I, great decoding potential can be found with $\rho_{\sigma,\mathbf{y}}(\Lambda)<1$.
Meanwhile, it seems that smaller sphere radius is required with the increment of SNR. This is because the received signal $\mathbf{y}$ is getting close to the lattice $\Lambda=\mathbf{Rx}$ as the effect of noises is suppressed gradually.


\subsection{LLL Reduction}
Lattice reduction techniques have a long tradition in the field of number theory.
In 1982, the celebrated LLL algorithm was proposed as a powerful and famous lattice reduction criterion for arbitrary lattice.
Specifically, a basis $\mathbf{B}$ is said to be LLL-reduced\footnote{Other
lattice reduction schemes like Korkin-Zolotarev (KZ) reduction and Seysen reduction also exist, see \cite{WubbenLRMagzine,Shanxiang1} for more details.}, if it satisfies the following two conditions:
\begin{itemize}
  \item $|\mu_{i,j}|\leq\frac{1}{2}$,\, for \, $1\leq j<i \leq n$,

  \item $\delta\|\mathbf{\widehat{b}}_i\|^{2} \leq \|\mu_{i+1,i} \mathbf{\widehat{b}}_i + \mathbf{\widehat{b}}_{i+1}\|^{2}$,\, for \, $1\leq i<n$,
\end{itemize}
where $\mathbf{\widehat{b}}_i$'s are the Gram-Schmidt vectors of the matrix $\mathbf{B}$ with $\min_{i}\|\widehat{\mathbf{b}}_i\|=\min_{i}|r_{i,i}|$ by QR-decomposition $\mathbf{B}=\mathbf{QR}$.
The first clause is called size reduction condition with $\mu_{i,j}=\langle\mathbf{b}_i,\mathbf{\widehat{b}}_j\rangle/\langle\mathbf{\widehat{b}}_j,\mathbf{\widehat{b}}_j\rangle$, while the second is known as Lov\'{a}sz condition. If Lov\'{a}sz condition is violated, the basis vectors $\mathbf{b}_i$ and $\mathbf{b}_{i+1}$ are swapped; otherwise, size reduction is carried out.
If only size reduction condition is satisfied, then the basis is called size-reduced.
The parameter $1/4 < \delta <1$ controls both the convergence speed of the reduction and the degree of orthogonality of the reduced basis.

Here, we highlight the significance of LLL reduction to the proposed ESD algorithm, which effectively improves $\min_{i}|r_{i,i}|$ (i.e., $\min_{i}\|\widehat{\mathbf{b}}_i\|$) through the matrix transformation (also reduce $\max_{i}|r_{i,i}|$ at the same time) \cite{LLLoriginal,CongProxity}.

Specifically, with LLL reduction, the system model in (\ref{eqn:QR}) is converted into an equivalent one, i.e.,
\begin{equation}
\mathbf{y}=\overline{\mathbf{R}}\mathbf{z}+\mathbf{n},
\label{nh1}
\end{equation}
where the LLL reduced matrix $\overline{\mathbf{R}}=\mathbf{RU}$ is more orthogonal than $\mathbf{R}$ with the unimodular matrix $\mathbf{U}\in\mathbf{R}^{n\times n}$ and $\mathbf{z}=\mathbf{U}^{-1}\mathbf{x}\in\mathbb{Z}^n$.
Then, according to (\ref{mp11}) in Theorem 1 and (\ref{mp1}) in Theorem 4, we can easily arrive at the following result.
\begin{my7}
To solve the same ILS problem, the proposed ESD algorithm with $\sigma=\min_{i}|\overline{r}_{i,i}|/(2\sqrt{\pi})$ given $\mathbf{y}=\overline{\mathbf{R}}\mathbf{z}+\mathbf{n}$ achieves a larger sphere radius than that with $\sigma=\min_{i}|r_{i,i}|/(2\sqrt{\pi})$ given $\mathbf{y}=\mathbf{R}\mathbf{x}+\mathbf{n}$ due to
\end{my7}
\vspace{-1em}
\begin{equation}
\min_{i}|\overline{r}_{i,i}|\geq\min_{i}|r_{i,i}|.
\end{equation}

Although LLL reduction is applied to increase the sphere radius, it is easy to check that the complexity by means of the number of visited nodes in ESD still obeys the upper bound $|S|<nK$.
Similarly, the upper bound $|L|<K$ for the number of collected candidate vectors holds as well, thus leading to a better decoding trade-off between performance and complexity.
On the other hand, the computational complexity of LLL reduction is known as $O(n^4\log n)$ while could be further reduced as $O(n^3\log n)$ through the effective LLL algorithm in \cite{CongEffectiveJJ}.
Furthermore, the complex LLL strategy can be applied to reduce the complexity \cite{4787140}.

\subsection{Complexity Reduction in Implementation}
Throughout the context, the infinite state space $\mathbf{x}\in\mathbb{Z}^n$ is considered, which means sufficient candidate nodes $\widehat{x}^j_i$ with $j=1,2,3, \ldots$ for each parent node $\underline{\widehat{x}^j_{i}}$ should be taken into account.
However, in practice, only limited candidate nodes need to be considered, and we now investigate the required size of index $j$.

From (\ref{fn2}), the normalized weighting of $j$th candidate node at searching layer $i$ can be written as
\begin{equation}
\hspace{-.1em}p(x^j_i)\hspace{-.2em}=\hspace{-.3em}\begin{cases}\hspace{-.2em}e^{-\frac{1}{2\sigma_i^2}((j-1)/2+d)^2}\hspace{-.6em}/\hspace{-.1em}\rho_{\sigma_i, \widetilde{x}_i}(\mathbb{Z})\ \text{when}\ j\ \text{is odd}, \\
       \hspace{-.2em}e^{-\frac{1}{2\sigma_i^2}(\frac{j}{2}-d)^2}\hspace{-.6em}/\hspace{-.1em}\rho_{\sigma_i, \widetilde{x}_i}(\mathbb{Z})\ \ \ \ \ \ \ \text{when}\ j\ \text{is even},
       \end{cases}
\label{pwq1}
\end{equation}
where $\frac{1}{2}\geq d=|x^1_i-\widetilde{x}_i|\geq0$.
Therefore, the summation normalized weighting of the first $2N$ candidate nodes with respect to $\widetilde{x}_i$ can be expressed as
\begin{equation}
P_{2N}\hspace{-.2em}=\hspace{-.2em}\sum^N_{j=1}\hspace{-.2em}\left(e^{-\frac{1}{2\sigma_i^2}(j-1+d)^2}\hspace{-.5em}+e^{-\frac{1}{2\sigma_i^2}(j-d)^2}\right)\hspace{-.2em}/\rho_{\sigma_i, \widetilde{x}_i}(\mathbb{Z}).
\end{equation}


Because of $\sum_{\widehat{x}^j_i\in\mathbb{Z}} p(\widehat{x}^j_i)=1$, with $\sigma=\min_{i}|r_{i,i}|/(2\sqrt{\pi})$ the normalized weighting (also can be viewed as probabilities in a one-dimensional distribution) except those $2N$ candidate nodes can be derived as
{\allowdisplaybreaks\begin{flalign}
1-P_{2N}&=\sum_{j\geq N+1}\hspace{-.4em}\left(e^{-\frac{1}{2\sigma_i^2}(j-1+d)^2}+e^{-\frac{1}{2\sigma_i^2}(j-d)^2}\right)\hspace{-.2em}/\rho_{\sigma_i, \widetilde{x}_i}(\mathbb{Z}) \notag\\
&<\sum_{j\geq N+1}\hspace{-.4em}2\cdot e^{-\frac{1}{2\sigma_i^2}(j-1)^2}\hspace{-.4em}/\rho_{\sigma_i, \widetilde{x}_i}(\mathbb{Z}) \notag\\
&<\sum_{j\geq N+1}2\cdot e^{-\frac{1}{2\sigma_i^2}[(j-1)^2-\frac{1}{4}]} /\rho_{\sigma_i}(\mathbb{Z})\notag\\
&\approx\sum_{j\geq N+1}2\cdot e^{-2\pi[(j-1)^2-\frac{1}{4}]} \notag\\
&=O\left(e^{-2\pi N^2}\right),
\label{xpq1}
\end{flalign}}which implies the tail bound (\ref{xpq1}) decays exponentially fast due to $e^{2\pi}\gg1$.

\begin{my7}
From (\ref{xpq1}), with $\sigma=\min_{i}|r_{i,i}|/(2\sqrt{\pi})$, only limited number of children candidate nodes are worthy being considered due to the negligible weighting $p(x^j_i), j>3$.
\end{my7}
Therefore, in practice, $j=3$ is recommended unless the initial searching size $K$ is sufficiently large.
This is also well suited to the practical cases for finite state space $\mathbf{x}\in\mathcal{X}^n$.
The same result about the choice of $j$ can also be derived through ESD with weighting $f(\cdot)$, which is omitted here due to simplicity.

\subsection{Optimization with respect to $\sigma$ by Feasible Relaxation }
As for the proposed ESD algorithm, the deviation factor $\sigma$ is fixed at $\min_{i}|r_{i,i}|/(2\sqrt{\pi})$, so that $K$ is adjustable to provide the tractable and flexible decoding trade-off.
However, the assumption $\mathbf{x}\in\mathbb{Z}^n$ may not hold in practice while decoding the ILS problem normally aims at a truncated state space of $\mathbf{x}\in\mathcal{X}^n$.
In this case, it is possible to further optimize $\sigma$ by this relaxation for a better decoding performance.

Specifically, let $\sigma=\frac{\min_{i}|r_{i,i}|}{\sqrt{2\log\alpha}}$ with $\alpha>1$. Then $\alpha$ becomes the parameter to be considered.
Moreover, with $\sigma=\frac{\min_{i}|r_{i,i}|}{\sqrt{2\log\alpha}}$, it has been demonstrated in \cite{Klein} that
\begin{equation}
\overset{n}{\underset{i=1}{\prod}}\rho_{\sigma_i, \widetilde{x}_i}(\mathbb{Z})\leq e^{\frac{2n}{\alpha}(1+O(\alpha^{-3}))},
\label{rho1}
\end{equation}
where the term $O(\alpha^{-3})$ in (\ref{rho1}) could be negligible if
$\alpha$ is large. Assume $\alpha$ satisfies this weak condition, by relaxation, (\ref{xnb1}) can be expressed as
\begin{equation}
e^{-\frac{2n}{\alpha}}\cdot\alpha^{-\|\mathbf{Rx}-\mathbf{y}\|^2/\text{min}_ir_{i,i}^2}\geq\frac{1}{K},
\end{equation}
which corresponds to
\begin{equation}
\|\mathbf{Rx}-\mathbf{y}\|\leq \text{min}_ir_{i,i}\cdot\sqrt{\text{log}_\alpha(Ke^{-2n/\alpha})}.
\label{distance2}
\end{equation}
Typically, this means candidate vectors $\widehat{\mathbf{x}}$'s with $\|\mathbf{Rx}-\mathbf{y}\|$ less than
the RHS of (\ref{distance2}) will be obtained by ESD.

In order to exploit the decoding potential, parameter $\alpha$ can be optimized to maximize the
upper bound shown in (\ref{distance2}). Hence, letting the derivative about
$\text{log}_\alpha(Ke^{-2n/\alpha})$ versus $\alpha$ be zero, the optimum $\alpha_{\text{o}}$
given the initial searching size $K$ can be determined by
\begin{equation}
K=(e\alpha_{\text{o}})^{2n/\alpha_{\text{o}}}.
\label{relationship K and rho}
\end{equation}
From (\ref{relationship K and rho}), it is easy to check that the optimum
$\alpha_{\text{o}}$ monotonically decreases with the increment of $K$,
which means the choice of $\sigma=\frac{\min_{i}|r_{i,i}|}{\sqrt{2\log\alpha_{\text{o}}}}$ should be improved with the increase of $K$ as well.
Note that such an optimization about $\sigma$ is only a compromise by relaxation, and $\sigma=\min_{i}|r_{i,i}|/(2\sqrt{\pi})$ is still a better choice for $\mathbf{x}\in\mathbb{Z}^n$.


\subsection{MMSE-based ESD}
In MIMO detection, the MMSE detector takes the signal-to-noise ratio (SNR) term (i.e., the SNR at each receive antenna is $1/\sigma_w^2$) into account and thereby leading to an improved performance. As shown in \cite{WubbenMMSE}, MMSE detector is equal to ZF (also known as Babai's rounding algorithm) with respect to an extended system model. To this end, we define the
$2n\times n$ extended channel matrix $\underline{\mathbf{B}}$ and the $2n\times1$ extended receive vector $\underline{\mathbf{c}}$:
\begin{equation}
\underline{\mathbf{B}}=\left[\begin{array}{c} \mathbf{B}\\ \sigma_w\mathbf{I}_n \end{array}\right]\ \ \text{and}\ \ \underline{\mathbf{c}}=\left[\begin{array}{c} \mathbf{c}\\ \mathbf{0}_{n,1} \end{array}\right]
\end{equation}
where $\mathbf{I}_n\in\mathbb{R}^{n\times n}$ is the identity matrix and $\mathbf{0}_{n,1}\in\mathbb{R}^{n\times1}$ is the zero vector.

This viewpoint allows us to incorporate the MMSE criterion into ESD to improve the decoding performance. Overall, the updated ESD algorithm is presented in Algorithm 2, where mechanisms of normalizing weighting, candidate protection, LLL reduction and so on are all considered for a better decoding trade-off.

\floatname{algorithm}{Algorithm}

\setcounter{algorithm}{1}

\begin{algorithm}[t]
\caption{Updated ESD Algorithm}
\begin{algorithmic}[1]
\Require
$K, \mathbf{R}, \mathbf{y}, L=\emptyset$
\Ensure
$\mathbf{Rx}\in\Lambda$
\State calculate $\alpha_o$ to obtain the optimized $\sigma$ according to (\ref{relationship K and rho})
\State by LLL reduction, transfer the system model to $\mathbf{y}\hspace{-.2em}=\hspace{-.2em}\overline{\mathbf{R}}\mathbf{z}+\mathbf{n}$
\State invoke \textbf{Function 2} with $i=n$ to search layer by layer
\State add all the candidates $\widehat{\mathbf{z}}$'s generated by \textbf{Function 2} to $L$
\State refresh the set $L$ by $\mathbf{x}=\mathbf{Uz}$
\State output $\widehat{\mathbf{x}}=\underset{\mathbf{x}\in L}{\operatorname{arg~min}} \, \|\mathbf{y}-\mathbf{R}\mathbf{x}\|$ as the decoding solution
\end{algorithmic}
\end{algorithm}

\floatname{algorithm}{Function}
%
\setcounter{algorithm}{1}

\begin{algorithm}[t]
\caption{Searching at layer $i$ given $[\widehat{z}_n,\ldots,\widehat{z}_{i+1}]$}
\begin{algorithmic}[1]
\State compute $\widetilde{z}_i$ according to (\ref{eqn:noise effection in determination})
\State compute $p(\widehat{z}^j_i)$ by (\ref{fn2}) with $j\in[1,2,3]$
\State compute $K(\widehat{z}^j_i)$ according to (\ref{prunethresholdl2})
\For {each specific integer candidate $\widehat{z}^j_i$}
\If {$K(\widehat{z}^j_i)<1$}
\State \hspace{-2em}prune $\widehat{z}^j_i$ from the tree searching
\Else
\State {\hspace{-2em}save $\widehat{z}^j_i$} to form the decoding result $[\widehat{z}_n,\ldots,\widehat{z}_{i+1}, \widehat{z}^j_i]$
\If {$2> K(\widehat{z}^j_i)\geq 1$}
\State \hspace{-2.5em}decode the rest of layers by SIC to get a candidate $\widehat{\mathbf{z}}$
\ElsIf {$K(\widehat{z}^j_i)\geq2$}
\If {$i=1$}
\State \hspace{-2em}output the candidate $\widehat{\mathbf{z}}$
\Else
\State \hspace{-2em}invoke \textbf{Function 2} to decode the next layer $i-1$
\EndIf
\EndIf
\EndIf
\EndFor
\end{algorithmic}
\end{algorithm}

\subsection{Soft-output Decoding}
Besides MIMO detection, the proposed ESD algorithm is also well suited for the soft-output detection in MIMO systems, which improves the performance by iteratively exchanging the extrinsic information between MIMO detector and soft-in soft-out (SISO) decoder.

Specifically, the extrinsic information in soft-output decoding is always calculated through the \emph{posterior} LLR for each information bit
associated with the transmitted signal $\mathbf{x}$ \cite{Hochwald,LarssonPartialMarginalization}.
For bit $b_i\in\{0,1\}$, the approximated LLR is computed as
\begin{equation}
L(b_i|\mathbf{c})=\text{log}\frac{\sum_{\mathbf{x}:b_{i}(\mathbf{x})=1}\text{exp}\ (-\frac{1}{2\sigma^2}\parallel \mathbf{c}-\mathbf{B}\mathbf{x} \parallel^2)}{\sum_{\mathbf{x}:b_{i}(\mathbf{x})=0}\text{exp}\ (-\frac{1}{2\sigma^2}\parallel \mathbf{c}-\mathbf{B}\mathbf{x} \parallel^2)},
\label{eqn:LLR}
\end{equation}
where $b_{i}(\mathbf{x})$ is the $i$th information bit associated with the obtained $\widehat{\mathbf{x}}$.
In this condition, ESD can be used to provide a set of collected candidate vectors (i.e., $L$) for the LLR computing.



\section{Simulation}
In this section, the performance and the complexity of the proposed ESD algorithm are evaluated by the large-scale MIMO detection.
Specifically, given the system model in (\ref{eqn:System Model}),
the $i$th entry of the transmitted signal $\mathbf{x}$, denoted as $x_i$, is a modulation symbol taken independently from an $M$-QAM constellation $\mathcal{X}$
with Gray mapping. Meanwhile, we assume a flat fading environment, where the square channel matrix
$\mathbf{B}$ contains uncorrelated complex Gaussian fading gains with unit
variance and remains constant over each frame duration. Let $E_b$ represent the average power per bit at the receiver. Then the signal-to-noise ratio (SNR) $E_b/N_0=n/(\text{log}_2(M)\sigma_w^2)$ where $M$ is the modulation level and $\sigma_w^2$ is the noise variance.
In particular, the ESD and the updated ESD algorithms described in this section are ESD with weighting $f(\cdot)$ and normalized weighting $p(\cdot)$ respectively.
Besides, both of them are enhanced by candidate protection, LLL reduction, MMSE augmentation as well as the optimized $\sigma$ through $\alpha_{\text{o}}$ in (\ref{relationship K and rho}).
As a fair comparison, all the other decoding schemes applied here are also strengthened by LLL reduction. Meanwhile, the sampling decoding schemes are also enhanced by MMSE augmentation.

\begin{figure}[t]
\includegraphics[width=3.5in]{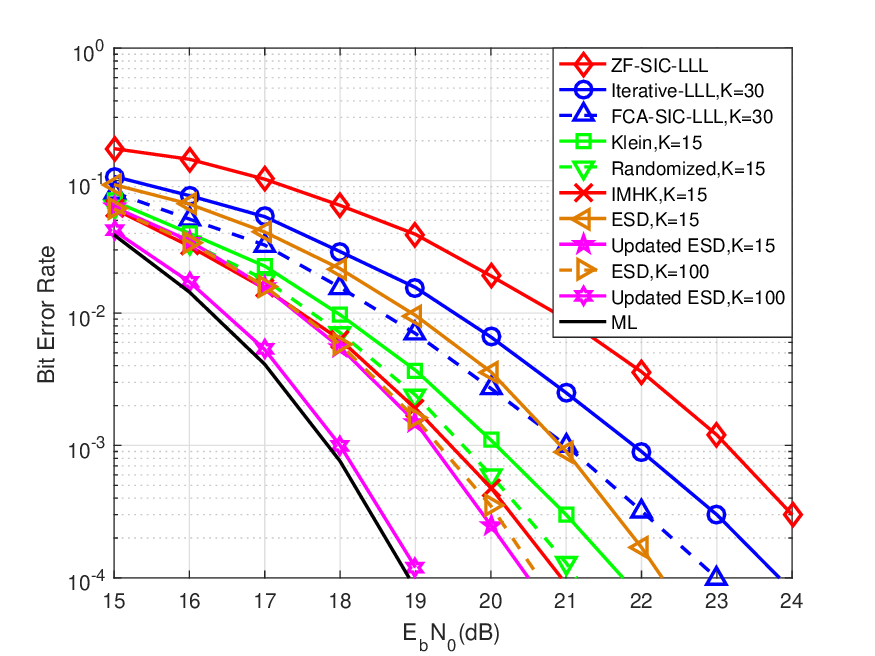}
\vspace{-1em}
  \caption{Bit error rate versus average SNR per bit for the uncoded $12 \times 12$ MIMO system using 64-QAM.}
  \label{simulation 2a}
\end{figure}


\begin{figure}[t]
\includegraphics[width=3.5in]{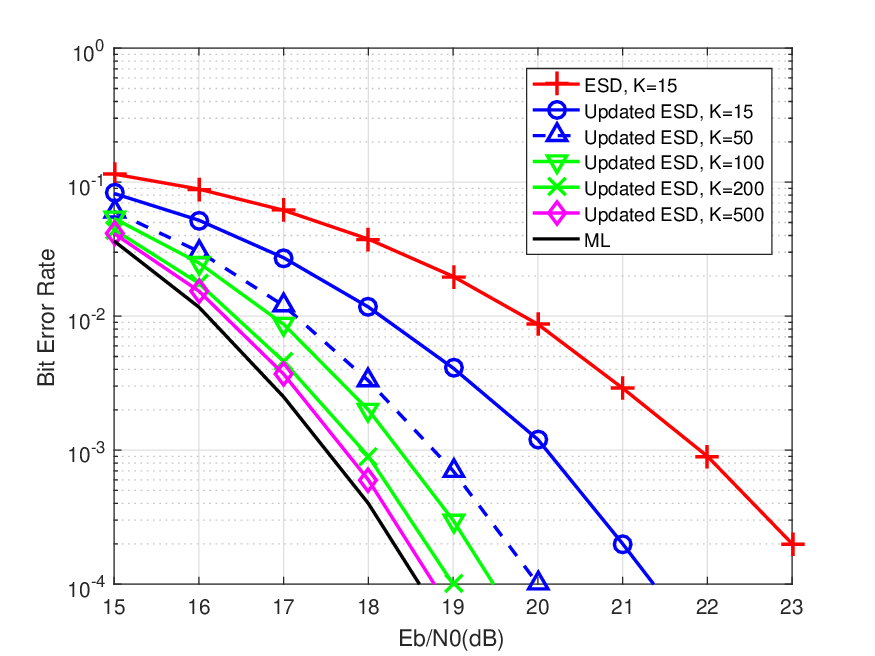}
\vspace{-1em}
  \caption{Bit error rate versus average SNR per bit for the uncoded $16 \times 16$ MIMO system using 64-QAM.}
  \label{simulation 2b}
\end{figure}

Fig. \ref{simulation 2a} shows the bit error rate (BER) of the proposed ESD algorithm compared with other decoding schemes in a $12\times12$ uncoded MIMO system with 64-QAM. Here, lattice-reduction-aided SIC (i.e., Babai's nearest plane) decoding serves as a performance baseline while ML decoding is implemented by the Schnorr-Euchner (SE) strategy from \cite{DamenDetectionSearch}. Clearly, compared to fixed candidates algorithm (FCA) in \cite{MaxiaoliSoft} and iterative list decoding in \cite{Shimokawa} with 30 samples, sampling decoding algorithms such as Klein's sampling decoding \cite{Klein}, randomized sampling decoding \cite{CongRandom} and IMHK sampling decoding \cite{ZhengWangTIT17} offer not only the improved BER performance but also the promise of smaller sample size $K$.
As for the proposed ESD algorithm, it is clear to see that ESD with weighting $f(\cdot)$ is not as good as sampling decoding under the same $K$. This is mainly because the initial searching size $K$ shrinks rapidly by $f(\cdot)$ especially for the limited state space $\mathbf{x}\in\mathcal{X}^n$ so that the decoding potential is not well exploited. Nevertheless, a decoding trade-off is still established by ESD with $f(\cdot)$, and one can improve the decoding performance by increasing $K$. As can be seen, there is a remarkable performance gain (i.e., near 2 dB) of ESD with $K=100$ over that with $K=15$.
However, since ESD with $f(\cdot)$ is essentially the same as Fincke-Pohst SD, it also implies the decoding trade-off of the conventional sphere decoding is not that charming.
On the other hand, as for the updated ESD, i.e., ESD with normalized weighting $p(\cdot)$, substantial performance gain can be found compared to ESD with $f(\cdot)$.
Besides, it is clear that the updated ESD outperforms all the sampling decoding schemes under the same size of $K$, which obeys the results shown in Table I.
More importantly, the complexity cost of the updated ESD is less than those of sampling decoding schemes, which is illustrated in Fig. 11 and Fig. 12 in detail.
Observe that with $K=100$, the performance of the updated ESD suffers negligible loss compared with ML. Therefore, with a moderate $K$, near-ML performance can be achieved.





\begin{figure}[t]
\includegraphics[width=3.5in]{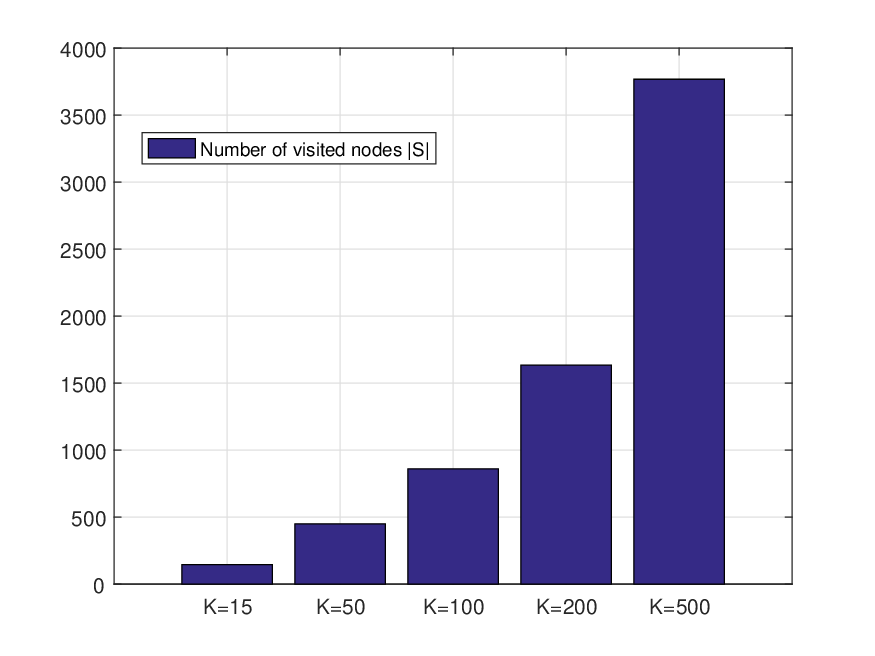}
\vspace{-1em}
  \caption{Number of visited nodes $|S|$ versus initial searching size $K$ for $16\times16$ uncoded MIMO using 64-QAM at SNR per bit = 17dB.}
  \label{simulation 3a}
\end{figure}

\begin{figure}[t]
\includegraphics[width=3.5in]{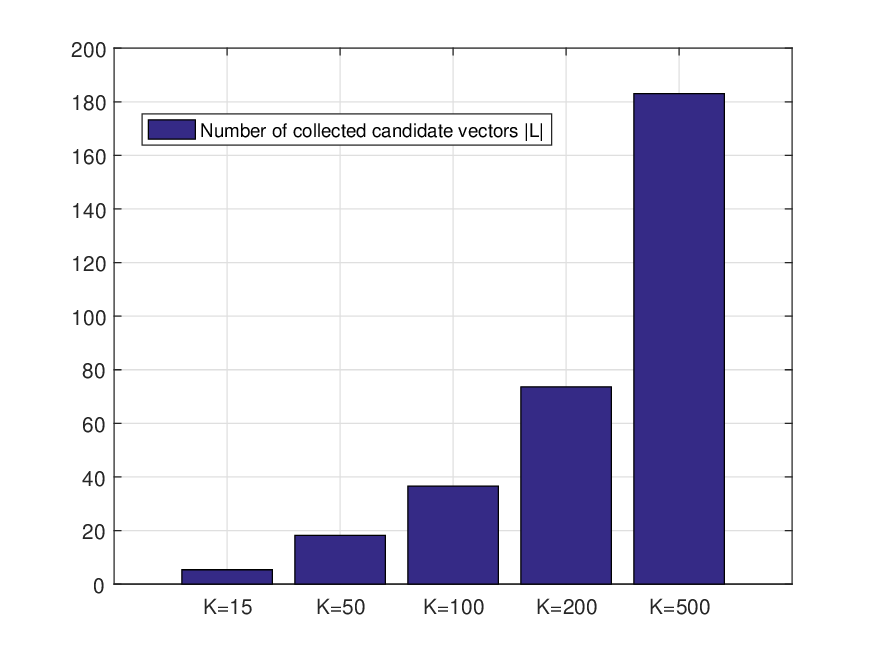}
\vspace{-1em}
  \caption{Number of collected candidate vectors $|L|$ versus initial searching size $K$ for $16\times16$ uncoded MIMO using 64-QAM at SNR per bit = 17dB.}
  \label{simulation 4a}
\end{figure}

\begin{figure}[t]
\includegraphics[width=3.5in]{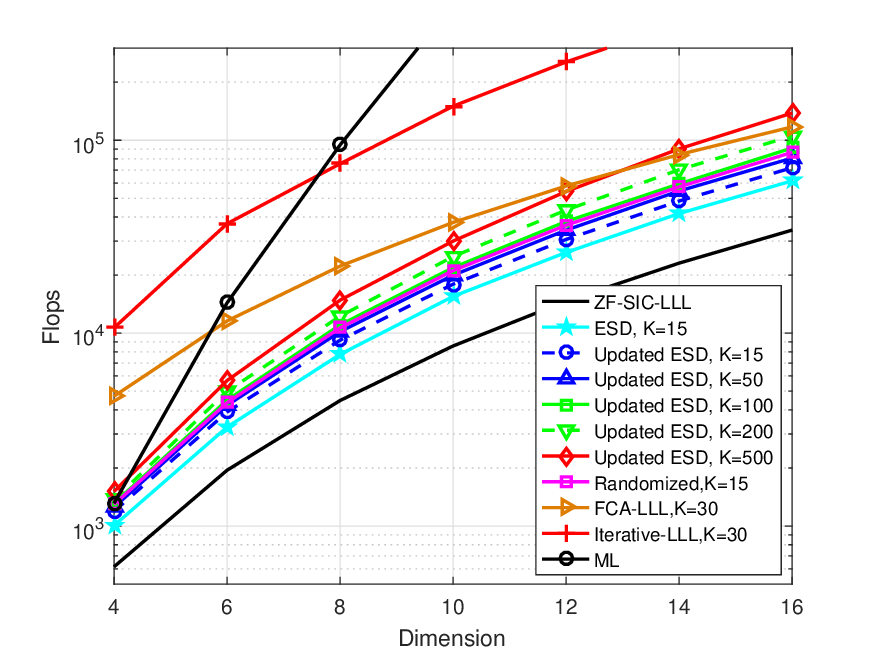}
\vspace{-1em}
  \caption{Complexity comparison in flops for the uncoded MIMO system using 64-QAM at SNR per bit = 17dB.}
  \label{simulation 2}
\end{figure}

\begin{figure}[t]
\includegraphics[width=3.5in]{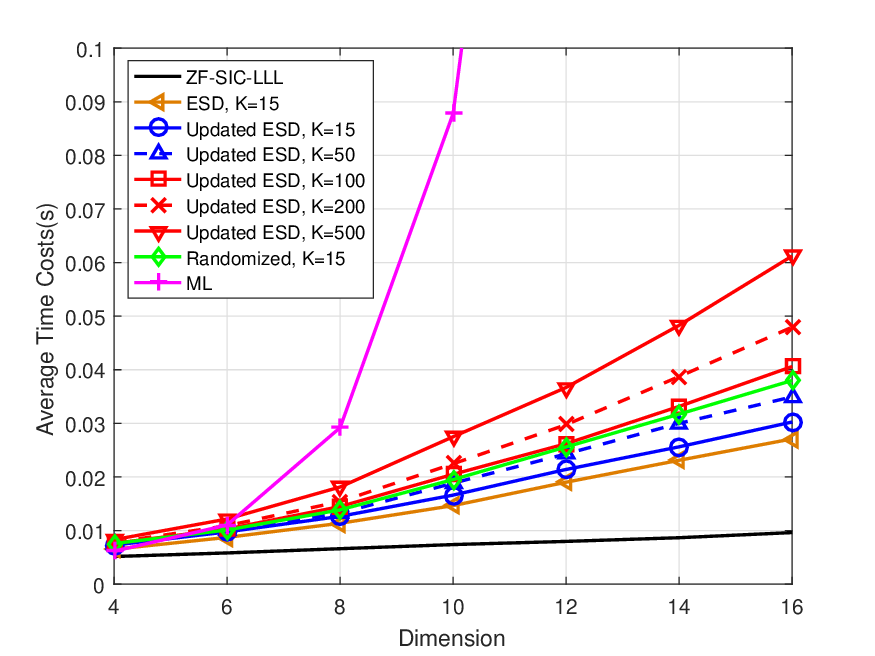}
\vspace{-1em}
  \caption{Complexity comparison in average time cost for the uncoded MIMO system using 64-QAM at SNR per bit = 17dB.}
  \label{simulation 3}
\end{figure}

In order to show the performance comparison with different initial searching sizes $K$, Fig. \ref{simulation 2b} is given to illustrate the BER performance of the proposed ESD algorithm in a $16\times16$ uncoded system with 64-QAM. According to (\ref{mp1}) in Theorem 4, a larger $K$ leads to a larger sphere radius $D$, which corresponds to a better decoding performance. More specifically, as shown in (\ref{prunethreshold22}), a larger $K$ naturally corresponds to a looser pruning threshold, which allows more candidate vectors to be obtained. Therefore, as can be seen clearly, with the increment of $K$, the BER performance improves gradually to the ML decoding performance. It is interesting to see that in Fig. \ref{simulation 2a} near-ML decoding performance can be achieved with $K=100$ while in Fig. \ref{simulation 2b} near-ML decoding performance requires $K=500$. This is because the larger system dimension has a deeper tree-structure to search, which requires more initial searching size $K$ to explore. Note that according to Theorem 7, the number of visited nodes and the number of collected candidate vectors are upper bounded by $|S|<nK$ and $|L|<K$ respectively, and the complexity increment with respect to $K$ is mild as expected, thus resulting in a promising trade-off between performance and complexity.


In Fig. 9 and Fig. 10, the comparisons about the average numbers of visited nodes number $|S|$ and collected candidate vectors $|L|$ obtained by the updated ESD for $16\times16$ uncoded MIMO systems using 64-QAM are given respectively.
Note that the $16\times16$ uncoded MIMO detection corresponds to the ILS problem with dimension $n = 32$.
Specifically, with the increment of $K$, both $|S|$ and $|L|$ improve gradually as more qualified candidate vectors are obtained by pruning threshold and candidate protection.
Clearly, even with the optimized $\sigma$ by relaxation, both $|S|$ and $|L|$ are always much smaller than the $nK$ and $K$ respectively.
This means the given upper bounds for $|S|$ and $|L|$ could be greatly refined, which will be one of our work in future.


Fig. \ref{simulation 2} shows the complexity comparison in flops of the proposed ESD algorithm with other decoding schemes in different system dimensions, where the flops evaluation scenario that we use comes from \cite{MatlabFlops}. Clearly, in the uncoded MIMO system with 64-QAM, ESD and updated ESD need much lower flops than other decoding schemes under the same size $K$. This benefit comes from the adaptation of the tree-structure searching, which reduces the computation in sampling procedures by removing all the unnecessary repetitions and calculations.
Specifically, the flops cost of the updated ESD with $K=50$ is less than that of randomized sampling decoding with $K=15$.
More importantly, with the increase of $K$, the decoding performance improves gradually but the complexity increment is mild.
Consequently, better BER performance and less complexity requirement make the proposed ESD algorithm very promising for solving the ILS problem in  large-scale MIMO detection.


Following the same scenario in Fig. \ref{simulation 2}, as a complement to illustrate the computational cost, Fig. \ref{simulation 3} is given to show the complexity comparison in average elapsed running times.
In particular, the uncoded MIMO system takes 64-QAM at SNR per bit = 17dB, and the simulation is conducted by MATLAB R2019a on a single computer, with an Intel Core i7 processor at 2.7GHz, a RAM of 8GB and Windows 10 Enterprise Service Pack operating system.
As can be seen clearly, the average elapsed running time of SIC-LLL decoding scheme increases slightly with the increase of system dimension. On the contrary, the optimal ML decoding from \cite{DamenDetectionSearch} takes an exponentially increasing average elapsed running time, which is unaffordable in most of cases. As expected, under the same $K$, the proposed ESD algorithm has a lower average elapsed running time than randomized sampling decoding, making it easy to be implemented especially in high-dimensional MIMO systems.


\section{Conclusions}
In this paper, extra degrees of freedom are introduced to sphere decoding for solving the ILS problem in large-scale MIMO detection.
Different from the conventional SD, the sphere radius of the proposed ESD algorithm is characterized by the initial searching size $K$ and the deviation factor $\sigma$. Based on it, we showed that the proposed ESD algorithm is exactly the same as the classic Fincke-Pohst SD but with a tractable decoding trade-off between performance and complexity. Moreover, to further exploit the decoding potential, two enhancement mechanisms named as normalized weighting and candidate protection are developed, which not only leads to a better decoding trade-off but also bridges the suboptimal and the optimal decoding performance by simply tuning the initial searching size $K\geq1$ freely.
In addition, further performance enhancement and complexity reduction are investigated to make the proposed ESD algorithm well suited to the various decoding requirements in large-scale MIMO detection.

%
%


\appendices

\bibliographystyle{IEEEtran}
\bibliography{IEEEabrv,reference1}

\end{document}